\documentclass[11pt,a4paper]{article}
\pdfoutput=1

\usepackage{jheppub}
\usepackage{amsmath}
\usepackage{amssymb}
\usepackage{graphics}
\usepackage[active]{srcltx}
\usepackage{pdfsync}
\usepackage{shuffle}
\usepackage{slashed}
\usepackage{hyperref}
\usepackage{subfigure}

\newcommand{\bea}{\begin{eqnarray}}
\newcommand{\eea}{\end{eqnarray}}
\newcommand{\nn}{\nonumber \\}

\def\W #1{\widetilde{#1}}

\def\braket#1{\left\langle #1 \right\rangle}

\def\Tr{\mathop{\rm Tr}}

\def\eref#1{(\ref{#1})}

\def\a{{\alpha}}

\def\b{{\beta}}

\def\vev{\braket}

\def\Spaa{\vev}

\allowdisplaybreaks

\title{Constructing tree amplitudes of scalar EFT from double soft theorem}
 \author[a]{Kang Zhou}

\affiliation[a]{Center for Gravitation and Cosmology, College of Physical Science and Technology, Yangzhou University,\\
No.180, Siwangting Road, Yangzhou, 225009, P.R. China.}

\emailAdd{zhoukang@yzu.edu.cn}

\date{\today}
\abstract{The well known Adler zero can fully determine tree amplitudes of non-linear sigma model (NLSM), but fails to fix tree pion amplitudes
with higher-derivative interactions. To fill this gap, in this paper we propose a new method based on exploiting the double soft theorem for scalars, which can
be applied to a wider range. A remarkable feature of this method is, we only assume the universality of soft behavior at the beginning, and determine the explicit form of double soft factor in the process of constructing amplitudes.
To test the applicability, we use this method to construct tree NLSM amplitudes and tree amplitudes those
pions in NLSM couple to bi-adjoint scalars. We also construct the simplest pion amplitudes which receive leading higher-derivative correction, with arbitrary number of external legs.
All resulted amplitudes are formulated as universal expansions to appropriate basis.
}

\keywords{Scattering Amplitudes, Soft Theorem}

\begin{document}

\maketitle \flushbottom

%%%%%%%%%%%%%%%%%%%%%%%%%%%%%%%%%%%%%%
\section{Introduction}
\label{sec-intro}
%%%%%%%%%%%%%%%%%%%%%%%%%%%%%%

In recent years, tree amplitudes of effective field theories (EFTs), which play an important and ubiquitous role in various branches of physics, have received remarkable attentions \cite{Kampf:2013vha,Cheung:2014dqa,Cheung:2015ota,He:2016iqi,Cheung:2016drk,Cheung:2018oki,Elvang:2018dco,Kampf:2019mcd,Bern:2021ppb,
AccettulliHuber:2021uoa,Carrasco:2021ptp,Brown:2023srz,Bonnefoy:2023imz,Bartsch:2024ofb}. Among these investigations, the study on tree amplitudes of nonlinear sigma model (NLSM) is an especially active area of research. This model describes the leading order dynamics of Nambu-Goldstone bosons in the low energy QCD, arises form the spontaneously breaking
of symmetry \cite{Brauner:2024juy}. The full Lagrangian for such Nambu-Goldstone bosons (pions) can be organized in a derivative expansion in the context of Chiral Perturbation Theory ($\chi^{\rm PT}$) as follows \cite{Weinberg:1978kz,Gasser:1983yg,Gasser:1984gg}
\bea
{\cal L}_{\chi^{\rm PT}}=\sum_{m=1}^\infty\,{\cal L}_{2m}\,,
\eea
where each ${\cal L}_{2m}$ contains operators with $2m$ derivatives. The the leading order Lagrangian ${\cal L}_2$ with $m=1$ is known as NLSM, while higher-order corrections correspond to $m\geq2$.
Tree amplitudes of the above model vanish in the soft limit, at any order $m$, due to the sift symmetry of Lagrangian.
This behavior is called Adler zero.

One of central aims of the modern S-matrix program is to build scattering amplitudes via a bottom-up method, free from reliance on a traditional Lagrangian or Feynman rules. The universal soft behavior of amplitudes is one appropriate physical criteria for realizing this goal, as studied in \cite{Nguyen:2009jk,Boucher-Veronneau:2011rwd,Rodina:2018pcb,Ma:2022qja,Cheung:2014dqa,Cheung:2015ota,Luo:2015tat,Elvang:2018dco}. It turns out that the tree NLSM amplitudes, which serve as the leading contribution to the full pion scattering at tree level, can be completely
determined by exploiting Adler zero \cite{Cheung:2014dqa,Cheung:2015ota}. However, Adler zero is not sufficient to fix higher-order amplitudes with higher-derivative corrections.

To fill this gap, the first attempt is made in \cite{Brown:2023srz}, by imposing the Bern-Carrasco-Johansson (BCJ)
relations \cite{Kawai:1985xq,Bern:2008qj,Chiodaroli:2014xia,Johansson:2015oia,Johansson:2019dnu} as physical constraints, instead of the weaker Adler zero.
In this paper, we propose another new approach from totally different point of view, based on inverting the double soft theorem.
Comparing with Adler zero, pion amplitudes do not vanish when two external particles are taken to be soft simultaneously, give rise to well defined
double soft theorem with universal soft factors, at leading and subleading orders \cite{Cachazo:2015ksa,Du:2015esa}. Obviously, such double soft behavior carries more
information than Adler zero, thus it is natural to expect that the technic of inverting double soft theorem has a wider range of applicability.

The method in this paper is the natural extension of the bottom-up approach developed in our recent works \cite{Zhou:2022orv,Wei:2023yfy,Hu:2023lso,Du:2024dwm}. In these previous works, we constructed Yang-Mills-scalar and Yang-Mills amplitudes by inverting single soft theorems, and represented the results with arbitrary number of external particles as the universal expansions to appropriate
basis. One significant feature of this method is, the explicit forms of soft factors are not assumed at the beginning. Oppositely, they are determined in the process of constructing amplitudes. In this paper, we generalize this approach to incorporate the double soft theorem.

In this paper, we restrict ourselves to the verification of applicability. Thus, we reconstruct ordered tree NLSM and ${\rm NLSM}\oplus{\rm BAS}$ amplitudes via our new
method, where BAS denotes the bi-adjoint-scalar theory, and also construct new ordered tree pion amplitudes with leading higher-derivative correction, valid for arbitrary number of external legs. The evaluation of more general higher-dimensional pion amplitudes is left to the future work. Although the double soft factors of NLSM amplitudes can be found in literatures \cite{Cachazo:2015ksa,Du:2015esa}, to make our method to be self-contained, we rederive them by using the technic similar as that in our previous work \cite{Du:2024dwm}. The plane is outlined as follows.

We first bootstrap the lowest $4$-point NLSM amplitudes, and represent them as expansions to bi-adjoint-scalar (BAS) amplitudes. Each $4$-point NLSM amplitude can be reinterpreted as the equivalent ${\rm NLSM}\oplus{\rm BAS}$ amplitude with two external pions and two external BAS scalars. Then, we generate ${\rm NLSM}\oplus{\rm BAS}$ amplitudes with two external pions and $n$ external scalars from these $4$-point ones, by inverting the single soft theorem
for BAS scalars. This is the key step, since we can derive the double soft factors from such $(n+2)$-point ${\rm NLSM}\oplus{\rm BAS}$ amplitudes.
By assuming the universality of double soft behavior, we invert the double soft theorem to construct ${\rm NLSM}\oplus{\rm BAS}$ amplitudes with
more external pions, as well as the pure NLSM amplitudes. The resulted amplitudes are automatically represented as universal expansions to ${\rm NLSM}\oplus{\rm BAS}$ amplitudes.

To achieve the pion amplitudes with leading higher-derivative correction, we first bootstrap the $4$-point amplitudes with higher mass dimension. At this step we employ the technic in the BCJ bootstrap method \cite{Brown:2023srz}, search for $4$-point solution which has correct mass dimension and satisfies BCJ relations.
It turns out that there is no consistent solution with mass dimension ${\cal D}+2$, and the unique solution with mass dimension ${\cal D}+4$ can be found, where ${\cal D}$ encodes the mass dimension of ordinary NLSM amplitudes. By inverting double soft theorem for pions, we then construct general amplitudes with mass dimension ${\cal D}+4$ from the $4$-point ones. The resulted amplitudes are again represented as universal expansions to ${\rm NLSM}\oplus{\rm BAS}$ amplitudes, and can be naturally interpreted as pion amplitudes
with a single insertion of higher-derivative local operator with lowest mass dimension.

Throughout our construction, the universality of soft behavior serves as the central priori assumption. On the other hand, the explicit form of double soft factor is derived, rather than assumed.
As mentioned before, all resulted amplitudes are expressed as expansions to ${\rm NLSM}\oplus{\rm BAS}$ amplitudes with less external pions and more external BAS scalars. Substituting the expansion of ${\rm NLSM}\oplus{\rm BAS}$ amplitudes iteratively leads to expansions to pure BAS basis. In above expansions, each coefficient solely depends on one of two orderings carried by the corresponding ${\rm NLSM}\oplus{\rm BAS}$ or BAS amplitude. This feature implies that the BCJ relations are automatically satisfied, since these relations hold for amplitudes in BAS basis.

The paper is organized as follows. In section \ref{sec-preparations}, we rapidly review the single- and double-soft behaviors of tree BAS amplitudes, which serve as the necessary preparation for subsequent works. In section \ref{sec-NLSM+BAS}, we construct tree ${\rm NLSM}\oplus{\rm BAS}$ amplitudes with two external pions and $n$ external BAS scalars, and derive the desired double soft theorem. By inverting double soft theorem, in section \ref{sec-constructNLSM} we construct general tree ${\rm NLSM}\oplus{\rm BAS}$ amplitudes, as well as tree NLSM amplitudes. Tree pion amplitudes with single insertion of higher-derivative operator are constructed in section \ref{sec-higherderi}. A brief discussion will be presented in section \ref{sec-summary}.

%%%%%%%%%%%%%%%%%%
\section{Brief review for single and double soft behaviors of BAS amplitudes}
\label{sec-preparations}
%%%%%%%%%%%%%%%%%%

The single and double soft behaviors of tree bi-adjoint scalar (BAS) amplitudes are essential for constructions in this paper. Thus, in this subsection, we give a brief review for such soft behaviors.

The BAS theory is the theory for massless scalar fields $\phi^{Aa}$ with cubic interaction, described by the Lagrangian
\bea
{\cal L}_{\rm BAS}={1\over2}\,\partial_\mu\phi^{Aa}\,\partial^{\mu}\phi^{Aa}+{\lambda\over3!}\,F^{ABC}f^{abc}\,
\phi^{Aa}\phi^{Bb}\phi^{Cc}\,,
\eea
where $F^{ABC}$ and $f^{abc}$ are structure constants of two Lie groups respectively.

The tree amplitudes of this theory only consists of propagators for massless scalars. Decomposing group factors gives
\bea
{\cal A}_n=\sum_{\sigma\in{\cal S}_n\setminus Z_n}\,\sum_{\sigma'\in{\cal S}'_n\setminus Z'_n}\,\Tr[T^{A_{\sigma_1}},\cdots T^{A_{\sigma_n}}]\,\Tr[T^{a_{\sigma'_1}}\cdots T^{a_{\sigma'_n}}]\,{\cal A}_{\rm BAS}(\sigma_1,\cdots,\sigma_n|\sigma'_1,\cdots,\sigma'_n)\,,
\eea
where ${\cal A}_n$ denotes the pure kinematic part of the full $n$-point amplitude with all coupling constants stripped off. The summation for $\sigma$ and $\sigma'$ are over all uncyclic permutations among external legs. Each partial amplitude ${\cal A}_{\rm BAS}(\sigma_1,\cdots,\sigma_n|\sigma'_1,\cdots,\sigma'_n)$ is simultaneously planar with respect to two orderings. In the reminder of this paper, we abbreviate ${\cal A}_{\rm BAS}(\sigma_1,\cdots,\sigma_n|\sigma'_1,\cdots,\sigma'_n)$ to ${\cal A}_{\rm BAS}(\vec{\pmb\sigma}_n|\vec{\pmb\sigma}'_n)$, where $\vec{\pmb\sigma}_n$ and $\vec{\pmb\sigma}'_n$ denote two ordered sets among $n$ elements. Some times we will fix one of two orderings without losing of generality, and explicitly write down the fixed one, for example ${\cal A}_{\rm BAS}(1,\cdots,n|\vec{\pmb\sigma}'_n)$.

The anti-symmetry of structure constants $F^{ABC}$ and $f^{abc}$ indicates that each partial amplitude carries an overall $\pm$ sign, arises from swapping lines at each vertex. We admit the convention that the overall sign is $+$ if the partial amplitude carries two same orderings. For instance, the amplitude ${\cal A}_{\rm BAS}(1,2,3,4|1,2,3,4)$ has the overall sign $+$. The sign for amplitudes which carry two different orderings can be generated from the above reference one by counting the number of flippings \cite{Cachazo:2013iea}.

For each double ordered partial BAS amplitude which solely consists of propagators, the leading single soft behavior obviously arises from the divergent propagators in the soft limit. Therefore, for $k_i^{\mu}\to \tau k_i^{\mu}$, $\tau\to 0$, the leading single soft theorem reads
\bea
{\cal A}^{(0)_i}_{\rm BAS}(1,\cdots,n|\vec{\pmb\sigma}_n)&=&S^{(0)_i}_s\,{\cal A}_{\rm BAS}(1,\cdots,i-1,i+1,\cdots,n|\vec{\pmb\sigma}_n\setminus i)\,,~~~\label{for-soft-fac-s}
\eea
where the soft factor $S^{(0)_i}_s$ is given by
\bea
S^{(0)_i}_s={1\over \tau}\,\left(\,{\delta_{i(i+1)}\over s_{i(i+1)}}+{\delta_{(i-1)i}\over s_{(i-1)i}}\,\right)\,.~~~~\label{soft-fac-s-0}
\eea
Here the symbol $\delta_{ij}$ is introduced to reflect the adjacency of external scalars $i$ and $j$ in the ordering $\vec{\pmb\sigma}_n$. If $i$, $j$ are not adjacent to each other, $\delta_{ij}=0$. If $i$ and $j$ are two adjacent elements, we have $\delta_{ij}=1$ when $i$ is on the l.h.s of $j$, and $\delta_{ij}=-1$ when $i$ is on the r.h.s of $j$. In \eref{soft-fac-s-0}, the Mandelstam variable $s_{i\cdots j}$ is defined as
\bea
s_{i\cdots j}\equiv k_{i\cdots j}^2\,,~~~{\rm with}~k_{i\cdots j}\equiv\sum_{a=i}^j\,k_a\,,~~~~\label{mandelstam}
\eea
where $k_a$ is the momentum carried by the external leg $a$. For $2$-particle channels contribute to the soft factor in \eref{soft-fac-s-0}, we have $s_{ij}=2k_i\cdot k_j$.

More generally, for BAS amplitude ${\cal A}_{\rm BAS}(\vec{\pmb\sigma}_n|\vec{{\pmb\sigma}}'_n)$ with two orderings denoted as $\vec{\pmb\sigma}_n$ and $\vec{\pmb\sigma}'_n$, one can introduce two symbols $\delta_{ij}$ and $\delta'_{ij}$ associated to $\vec{\pmb\sigma}_n$ and $\vec{\pmb\sigma}'_n$ respectively, in exactly the same way. Then the leading soft behavior is represented as
\bea
{\cal A}^{(0)_i}_{\rm BAS}(\vec{\pmb\sigma}_n|\vec{\pmb\sigma}'_n)&=&S^{(0)_i}_s\,{\cal A}_{\rm BAS}(\vec{\pmb\sigma}_n\setminus i|\vec{\pmb\sigma}'_n\setminus i)\,,~~~\label{soft-s1}
\eea
with the following soft factor
\bea
S^{(0)_i}_s={1\over\tau}\,\sum_{j\neq i}\,{\delta_{ij}\,\delta'_{ij}\over s_{ij}}\,.~~\label{soft-fac-s1}
\eea

When two legs $i$ and $j$ are taken to be soft simultaneously, the corresponding leading double soft behavior arises from Feynman diagrams those $i$
and $j$ couple to a common vertex, then couple to another external leg, namely,
\bea
{\cal A}^{(0)_{ij}}_{\rm BAS}(\vec{\pmb\sigma}_n|\vec{\pmb\sigma}'_n)&=&S^{(0)_{ij}}_s\,{\cal A}_{\rm BAS}(\vec{\pmb\sigma}_n\setminus \{ij\}|\vec{\pmb\sigma}'_n\setminus \{ij\})\,,~~~\label{soft-2s1}
\eea
with the soft factor
\bea
S^{(0)_{ij}}_s={1\over\tau^3}\,\sum_{h\neq i,j}\,{\delta_{h(ij)}\,\delta_{ij}\,\delta'_{h(ij)}\,\delta'_{ij}\over 4\,(k_i\cdot k_j)\,(k_{ij}\cdot k_h)}\,.~~\label{soft-fac-2s}
\eea
In \eref{soft-fac-2s}, we introduced the new symbol $\delta_{h(ij)}$ as follows. For $\delta_{ij}\neq0$, one can regard two legs $i$ and $j$ as an individual element $(i,j)$. Then, $\delta_{h(ij)}=0$ when $h$ is not adjacent to $(ij)$ in the ordering $\vec{\pmb\sigma}_n$. When $h$ is adjacent to $(ij)$, we have $\delta_{h(ij)}=+1$ if $h$ is on the l.h.s of $(ij)$, and $\delta_{h(ij)}=-1$ if $h$ is on the r.h.s. The definition of $\delta'_{h(ij)}$ which depends on the ordering $\vec{\pmb\sigma}'_n$ is analogous.

The antisymmetry of the symbol $\delta_{ij}$ leads to the following helpful property
\bea
& &\left({\delta_{ab}\over s_{ab}}+{\delta_{bc}\over s_{bc}}\right)\,{\cal A}_{\rm BAS}(1,\cdots,a,k_1,\cdots,k_p,c,\cdots,n|\vec{\pmb\sigma}_n\setminus b)\nn
&=&\left({\delta_{ab}\over s_{ab}}+{\delta_{bk_1}\over s_{bk_1}}\right)\,{\cal A}_{\rm BAS}(1,\cdots,a,k_1,\cdots,k_p,c,\cdots,n|\vec{\pmb\sigma}_n\setminus b)\nn
& &+\sum_{i=1}^{p-1}\,\left({\delta_{k_ib}\over s_{k_ib}}+{\delta_{bk_{i+1}}\over s_{bk_{i+1}}}\right)\,{\cal A}_{\rm BAS}(1,\cdots,a,k_1,\cdots,k_p,c,\cdots,n|\vec{\pmb\sigma}_n\setminus b)\nn
& &+\left({\delta_{k_pb}\over s_{k_pb}}+{\delta_{bc}\over s_{bc}}\right)\,{\cal A}_{\rm BAS}(1,\cdots,a,k_1,\cdots,k_p,c,\cdots,n|\vec{\pmb\sigma}_n\setminus b)\,.~~\label{tech}
\eea
This relation can be extended to the double soft case as follows,
\bea
& &\left({\delta_{a(bc)}\over 2k_{bc}\cdot k_a}+{\delta_{(bc)d}\over 2k_{bc}\cdot k_d}\right)\,{\cal A}_{\rm BAS}(1,\cdots,a,k_1,\cdots,k_p,d,\cdots,n|\vec{\pmb\sigma}_n\setminus \{b,c\})\nn
&=&\left({\delta_{a(bc)}\over 2k_{bc}\cdot k_a}+{\delta_{(bc)k_1}\over 2k_{bc}\cdot k_{k_1}}\right)\,{\cal A}_{\rm BAS}(1,\cdots,a,k_1,\cdots,k_p,d,\cdots,n|\vec{\pmb\sigma}_n\setminus \{b,c\})\nn
& &+\sum_{i=1}^{p-1}\,\left({\delta_{k_i(bc)}\over 2k_{bc}\cdot k_{k_i}}+{\delta_{(bc)k_{i+1}}\over 2k_{bc}\cdot k_{k_{i+1}}}\right)\,{\cal A}_{\rm BAS}(1,\cdots,a,k_1,\cdots,k_p,d,\cdots,n|\vec{\pmb\sigma}_n\setminus \{b,c\})\nn
& &+\left({\delta_{k_p(bc)}\over 2k_{bc}\cdot k_p}+{\delta_{(bc)d}\over 2k_{bc}\cdot k_d}\right)\,{\cal A}_{\rm BAS}(1,\cdots,a,k_1,\cdots,k_p,d,\cdots,n|\vec{\pmb\sigma}_n\setminus \{b,c\})\,.~~\label{tech-double}
\eea
The above decompositions will be used frequently in subsequent sections.

%%%%%%%%%%%%%%%%%%%%%%%%%%%%%%%%%%%%%
\section{Tree NLSM$\oplus$BAS amplitudes and double soft theorem}
\label{sec-NLSM+BAS}
%%%%%%%%%%%%%%%%%%%%%%%%%%%%%%%%%%%%%

This section aims to construct ${\rm NLSM}\oplus{\rm BAS}$ amplitudes with $2$ external pions
and $n$ external BAS scalars, and derive the subleading soft theorem. The ${\rm NLSM}\oplus{\rm BAS}$ amplitudes
under consideration are understood as pions in NLSM couple with BAS scalars. In subsection \ref{subsec-4pNLSM}, we bootstrap $4$-point
NLSM amplitudes by exploiting physical criteria like mass mass dimension and proper symmetry, and reinterpret them as the equivalent $4$-point
${\rm NLSM}\oplus{\rm BAS}$ amplitudes. Subsequently, we construct ${\rm NLSM}\oplus{\rm BAS}$ amplitudes with $2$ external pions
and $n$ external BAS scalars in subsection \ref{subsec-n+2pNLSM+BAS}, by inverting the leading soft theorem for BAS amplitudes. Then, in
subsection \ref{subsec-doublesoft}, we derive the desired double soft theorem from the $(n+2)$-point ${\rm NLSM}\oplus{\rm BAS}$ amplitudes obtained
in subsection \ref{subsec-n+2pNLSM+BAS}.

Although our method is exactly bottom-up,
for readers' convenience, it is worth to give the related Lagrangian here.
The standard NLSM Lagrangian in the Cayley parametrization is
\bea
{\cal L}_{\rm NLSM}={1\over 8\lambda^2}\,{\rm Tr}(\partial_\mu {\rm U}^\dag\partial^\mu {\rm U})\,,~~\label{Lag-N}
\eea
with
\bea
{\rm U}=(\mathbb{I}+\lambda\Phi)\,(\mathbb{I}-\lambda\Phi)^{-1}\,,
\eea
where $\mathbb{I}$ is the identity matrix, and $\Phi=\phi_IT^I$, with $T^I$ the generators of $U(N)$.
The Lagrangian \eref{Lag-N} indicates that the mass dimension of the kinematic part of any $n$-point NLSM amplitude is always ${\cal D}=2$.

%%%%%%%%%%%%%%%%%%%%%%%%%%%%%%%%%%%
\subsection{$4$-point NLSM amplitudes}
\label{subsec-4pNLSM}
%%%%%%%%%%%%%%%%%%%%%%%%%%%%%%%%%%

The $3$-point tree NLSM amplitude does not exist, due to the mass dimension, and the kinematical constraint $k_i\cdot k_j=0$ for any $i,j\in\{1,2,3\}$ based on momentum conservation and on-shell conditions.
The simplest tree NLSM amplitudes are $4$-point ones ${\cal A}_{\rm NLSM}(\vec{\pmb\sigma}_4)$. Such amplitudes can not factorize into lower-point ones, thus do not involve any pole. This observation, together with the correct mass dimension, require the $4$-point amplitudes to be linear combinations of Mandelstam variables $s$, $t$ and $u$, where $s=s_{12}$, $t=s_{14}$, $u=s_{13}$, satisfying $s+u+t=0$. Then each amplitude with particular ordering $\vec{\pmb\sigma}_4$ can be determined by considering the symmetry. For example, the ordering carried by ${\cal A}_{\rm NLSM}(1,2,3,4)$ indicates the symmetry among $s$ and $t$, thus ${\cal A}_{\rm NLSM}(1,2,3,4)$ is proportional to $s+t$ or $u$.
We can choose
\bea
{\cal A}_{\rm NLSM}(1,2,3,4)=s+t=-u\,,~~\label{4p1}
\eea
via an overall rescaling of amplitude. Similarly, we have
\bea
& &{\cal A}_{\rm NLSM}(1,3,2,4)=t+u=-s\,,\nn
& &{\cal A}_{\rm NLSM}(1,2,4,3)=u+s=-t\,.~~\label{4p2}
\eea
The cyclicity of orderings, and the ordered reversed relation ${\cal A}(a_1,\cdots,a_n)=(-)^n{\cal A}(a_n,\cdots,a_1)$ for arbitrary ordered amplitudes, indicate that $\langle1,2,3,4\rangle$, $\langle1,3,2,4\rangle$ and $\langle1,2,4,3\rangle$ in \eref{4p1} and \eref{4p2} cover all inequivalent orderings among $4$ elements. Here and after, we use $\Spaa{\cdots}$ and $\{\cdots\}$ to denote ordered and unordered sets, respectively. It is straightforward to verify that amplitudes in \eref{4p1} and \eref{4p2} satisfy both Kleiss-Kuijf (KK) \cite{Kleiss:1988ne} and BCJ relations for ordered amplitudes.

As well known, each tree amplitude for massless particles can be expanded to BAS amplitudes, with each coefficient the polynomial of Lorentz invariants depend on kinematic variables\footnote{A brief explanation can be seen in the Appendix in \cite{Du:2024dwm}}. The KK and BCJ relations among BAS amplitudes indicates that there is only one independent $4$-point BAS amplitude with fixed $\vec{\pmb\sigma}_4$. Thus, we expect that each $4$-point NLSM
amplitude ${\cal A}_{\rm NLSM}(\vec{\pmb\sigma}_4)$ satisfies
\bea
{\cal A}_{\rm NLSM}(\vec{\pmb\sigma}_4)&=&C(\vec{\pmb\sigma}_4)\,{\cal A}_{\rm BAS}(1,2,3,4|\vec{\pmb\sigma}_4)\,.~~\label{equ-c}
\eea
Evaluating ${\cal A}_{\rm BAS}(1,2,3,4|\vec{\pmb\sigma}_4)$ for inequivalent $\vec{\pmb\sigma}_4$ turns equations in \eref{equ-c} to
\bea
C(1,2,3,4)\,\Big({1\over s}+{1\over t}\Big)&=&s+t\,,\nn
C(1,3,2,4)\,\Big(-{1\over t}\Big)&=&-s\,,\nn
C(1,2,4,3)\,\Big(-{1\over s}\Big)&=&-t\,,
\eea
and the solution is found to be $C(1,2,3,4)=C(1,3,2,4)=C(1,2,4,3)=st$. This solution shows that the coefficients $C(\vec{\pmb\sigma}_4)$ are independent of the ordering $\vec{\pmb\sigma}_4$. From the Cachazo-He-Yuan (CHY) point of view, this character means the corresponding CHY integrand factorizes
as ${\cal I}_{\rm CHY}={\cal I}_L\,{\cal I}_R$, where ${\cal I}_R$ is the Parke-Taylor factor for the ordering $\vec{\pmb\sigma}_4$. Therefore,
the independence on $\vec{\pmb\sigma}_4$ is equivalent to the well known double copy structure.
For latter convenience, we express the coefficient $st$ as $st=4k_3\cdot P_2\cdot k_1$, where the tensor $P_i^{\mu\nu}$ is defined as $P_i^{\mu\nu}\equiv k^\mu_i k^\nu_i$. Consequently, the $4$-point NLSM amplitudes can be expressed as the unified formula
\bea
{\cal A}_{\rm NLSM}(\vec{\pmb\sigma}_4)=(k_3\cdot P_2\cdot k_1)\,{\cal A}_{\rm BAS}(1,2,3,4|\vec{\pmb\sigma}_4)\,,~~\label{4pNLSM-expan}
\eea
up to an overall rescaling.

We can reinterpret ${\cal A}_{\rm NLSM}(\vec{\pmb\sigma}_4)$ as the amplitude of ${\rm NLSM}\oplus{\rm BAS}$ theory with two external pions and two external BAS scalars, interact via a specific vertex which is equivalent to the quadrivalent operator in NLSM, based on the following discussion. A pion and a BAS scalar carry exactly the same mass and spin, the only difference is that each BAS scalar enters two orderings while each pion only enters one. However, for the special case that the amplitude includes two BAS scalars, the ordering among two BAS scalars is meaningless due to the cyclic invariance. Furthermore, let us
compare the expansion \eref{4pNLSM-expan} with the $4$-point Yang-Mills-scalar (YMS) amplitude \cite{Fu:2017uzt,Zhou:2022orv,Du:2024dwm}
\bea
{\cal A}_{\rm YMS}(1,4;\{2,3\}|\vec{\pmb\sigma}_4)&=&(\epsilon_3\cdot k_1)\,{\cal A}_{\rm YMS}(1,3,4;2|\vec{\pmb\sigma}_4)\nn
&=&(\epsilon_3\cdot f_2\cdot k_1)\,{\cal A}_{\rm BAS}(1,2,3,4|\vec{\pmb\sigma}_4)\,,~~\label{4p-YMS}
\eea
where $1$ and $4$ are two external BAS scalars while $\{2,3\}$ denotes the unordered set of external gluons $2$ and $3$.
Here $\epsilon_i$ are polarization vectors carried by gluon $i$, and the strength tensor are defined as $f_i^{\mu\nu}\equiv k_i^\mu\epsilon_i^\nu-\epsilon_i^\mu k_i^\nu$. Comparing expansions \eref{4pNLSM-expan} and \eref{4p-YMS}, we see that the $4$-point NLSM amplitude can be generated from the
$4$-point YMS amplitude ${\cal A}_{\rm YMS}(1,4;\{2,3\}|\vec{\pmb\sigma}_4)$ through the following specific dimensional reduction: consider the $d+1$-dimensional space-time, let $\epsilon_3$ and $\epsilon_2$ to lie in the extra dimension with $\epsilon_3\cdot\epsilon_2=-k_2\cdot k_3$, while all $k_i$ are kept to be $d$-dimensional. From the $d$-dimensional point of view, the above manipulation exactly reduces the expansion \eref{4p-YMS}
to the scalar amplitude \eref{4pNLSM-expan}. At the YMS side, external particles $1$ and $4$ are two BAS scalars, and the dimensional reduction described above does not affect these two particles.
Thus, it is natural to interpret $1$ and $4$ as BAS scalars in the resulted amplitude. Consequently, the expansion \eref{4pNLSM-expan} can also be understood as the $4$-point ${\rm NLSM}\oplus{\rm BAS}$ amplitude
\bea
{\cal A}_{{\rm N}\oplus{\rm B}}(1,2;\{a,b\}|\vec{\pmb\sigma}_4)=(k_b\cdot P_a\cdot k_1)\,{\cal A}_{\rm BAS}(1,a,b,2|\vec{\pmb\sigma}_4)\,,~~\label{4pN+B}
\eea
where $1$ and $2$ are two external BAS scalars, $a$ and $b$ are two external pions.

%%%%%%%%%%%%%%%%%%%%%%%%%%%%%%%%%%%%%
\subsection{$(n+2)$-point ${\rm NLSM}\oplus{\rm BAS}$ amplitudes with $2$ external pions}
\label{subsec-n+2pNLSM+BAS}
%%%%%%%%%%%%%%%%%%%%%%%%%%%%%%%%%%%%%%

Start from $4$-point ${\rm NLSM}\oplus{\rm BAS}$ amplitudes in \eref{4pN+B}, one can invert the leading soft theorem for BAS scalars to construct ${\rm NLSM}\oplus{\rm BAS}$ amplitudes with more external BAS scalars. The method is based on assuming the universality of soft theorem for BAS amplitudes, introduced in section \ref{sec-preparations}.

Consider the $5$-point ${\rm NLSM}\oplus{\rm BAS}$ amplitude ${\cal A}_{{\rm N}\oplus{\rm B}}(1,2,3;\{a,b\}|\vec{\pmb\sigma}_5)$, and take
$k_2\to\tau k_2$, $\tau\to 0$. The soft theorem in \eref{for-soft-fac-s} and \eref{soft-s1} forces the leading soft behavior to be
\bea
{\cal A}^{(0)_2}_{{\rm N}\oplus{\rm B}}(1,2,3;\{a,b\}|\vec{\pmb\sigma}_5)&=&{1\over\tau}\,\Big({\delta_{12}\over s_{12}}+{\delta_{23}\over s_{23}}\Big)\,{\cal A}(1,3;\{a,b\}|\vec{\pmb\sigma}_5\setminus2)\nn
&=&{1\over\tau}\,\Big({\delta_{12}\over s_{12}}+{\delta_{23}\over s_{23}}\Big)\,\Big((k_b\cdot P_a\cdot k_1)\,{\cal A}_{\rm BAS}(1,a,b,3|\vec{\pmb\sigma}_5\setminus2)\Big)\nn
&=&(k_b\cdot P_a\cdot k_1)\,{\cal A}^{(0)_2}_{\rm BAS}(1,\langle a,b\rangle\shuffle 2,3|\vec{\pmb\sigma}_5)\,,~~\label{5p-soft}
\eea
where the second equality uses the formula of $4$-point amplitudes in \eref{4pN+B}, and the third is obtained via the decomposition in \eref{tech}.
As mentioned in the previous subsection, the notation $\Spaa{\cdots}$ is used to distinguish ordered sets from unordered ones which are labeled by $\{\cdots\}$.
The shuffle $\vec{\pmb\a}\shuffle\vec{\pmb\b}$ means summing over all permutations such that the relative order in each of the ordered sets
$\vec{\pmb\a}$ and $\vec{\pmb\b}$ is kept. For instance, suppose $\vec{\pmb\a}=\Spaa{1,2}$, $\vec{\pmb\b}=\Spaa{3,4}$, then
$\vec{\pmb\a}\shuffle\vec{\pmb\b}$ is understood as
\bea
{\cal A}(\vec{\pmb\a}\shuffle\vec{\pmb\b})={\cal A}(1,2,3,4)+{\cal A}(1,3,2,4)+{\cal A}(1,3,4,2)\,.
\eea
In \eref{5p-soft}, the soft factor is governed by two orderings $\Spaa{1,2,3}$ and $\vec{\pmb\sigma}_5$, it follows from the universality
of soft behavior:
the soft factor in \eref{soft-fac-s1} carries two symbols $\delta_{ij}$ and
$\delta'_{ij}$, thus requires two corresponding orderings. Therefore, such soft operator never acts on external pions which belong to only one ordering.

The soft behavior in \eref{5p-soft} implies that the amplitude ${\cal A}^{(0)_2}_{{\rm N}\oplus{\rm B}}(1,2,3;\{a,b\}|\vec{\pmb\sigma}_5)$ can be expanded to BAS amplitudes as
\bea
{\cal A}_{{\rm N}\oplus{\rm B}}(1,2,3;\{a,b\}|\vec{\pmb\sigma}_5)=C(\shuffle)\,{\cal A}_{\rm BAS}(1,\langle a,b\rangle\shuffle 2,3|\vec{\pmb\sigma}_5)\,,~~\label{5p-2soft}
\eea
where the coefficients $C(\shuffle)$ satisfy
\bea
C(\shuffle)\Big|_{k_2\to\tau k_2,\tau\to 0}=k_b\cdot P_a\cdot k_1\,.~~\label{5p-coe}
\eea
To fix coefficients $C(\shuffle)$, we consider $k_1\to\tau k_1$ to get
\bea
{\cal A}^{(0)_1}_{{\rm N}\oplus{\rm B}}(1,2,3;\{a,b\}|\vec{\pmb\sigma}_5)&=&{1\over\tau}\,\Big({\delta_{31}\over s_{31}}+{\delta_{12}\over s_{12}}\Big)\,{\cal A}(2,3;\{a,b\}|\vec{\pmb\sigma}_5\setminus1)\nn
&=&{1\over\tau}\,\Big({\delta_{31}\over s_{31}}+{\delta_{12}\over s_{12}}\Big)\,\Big((k_b\cdot P_a\cdot k_2)\,{\cal A}_{\rm BAS}(2,a,b,3|\vec{\pmb\sigma}_5\setminus1)\Big)\nn
&=&(k_b\cdot P_a\cdot k_2)\,{\cal A}^{(0)_1}_{\rm BAS}(1,2,a,b,3|\vec{\pmb\sigma}_5)\,.~~\label{5p-1soft}
\eea
Comparing the soft behavior in \eref{5p-1soft} with \eref{5p-2soft} and \eref{5p-coe}, we find
\bea
C(\shuffle)=k_b\cdot P_a\cdot X_a\,,~~\label{C-result}
\eea
where each combinatorial momentum $X_a$ is defined as the summation of momenta carried by legs at the l.h.s of $a$ in the left ordering of the corresponding BAS amplitude. As can be straightforwardly verified, with coefficients $C(\shuffle)$ defined in \eref{C-result}, the expansion in \eref{5p-2soft} also gives correct soft behavior for $k_3\to\tau k_3$.
Consequently, the $5$-point amplitude ${\cal A}_{{\rm N}\oplus{\rm B}}(1,2,3;\{a,b\}|\vec{\pmb\sigma}_5)$ can be expanded as
\bea
{\cal A}_{{\rm N}\oplus{\rm B}}(1,2,3;\{a,b\}|\vec{\pmb\sigma}_5)=(k_b\cdot P_a\cdot X_a)\,{\cal A}_{\rm BAS}(1,\langle a,b\rangle\shuffle 2,3|\vec{\pmb\sigma}_5)\,.~~\label{5p-expan}
\eea

Repeating the above manipulation recursively, the $(n+2)$-point ${\rm NLSM}\oplus{\rm BAS}$ amplitudes with $n$ external BAS scalars and $2$ external pions are found to be
\bea
{\cal A}_{{\rm N}\oplus{\rm B}}(1,\cdots,n;\{a,b\}|\vec{\pmb\sigma}_{n+2})=(k_b\cdot P_a\cdot X_a)\,{\cal A}_{\rm BAS}(1,\langle a,b\rangle\shuffle \langle2,\cdots,n-1\rangle,n|\vec{\pmb\sigma}_{n+2})\,.~~\label{n+2p-expan}
\eea
In the above expansion formula \eref{n+2p-expan}, the permutation symmetry among pions $a$ and $b$ is not manifest. Such symmetry is protected by the relation
\bea
& &(k_a\cdot X_a)\,{\cal A}_{\rm BAS}
(1,\langle2,\cdots,n-1\rangle\shuffle\langle a,b\rangle,n|\vec{\pmb{\sigma}}_{n+2})\nn
&=&(k_b\cdot X_b)\,{\cal A}_{\rm BAS}
(1,\langle2,\cdots,n-1\rangle\shuffle\langle b,a\rangle,n|\vec{\pmb{\sigma}}_{n+2})\,,~~\label{rela}
\eea
which can be proved as follows. The fundamental BCJ relation indicates
\bea
& &(k_a\cdot X_a)\,{\cal A}_{\rm BAS}
(1,\Spaa{2,\cdots,n-1}\shuffle\Spaa{a,b},n|\vec{\pmb{\sigma}}_{n+2})\nn
&=&-(k_a\cdot X_a)\,{\cal A}_{\rm BAS}
(1,\Spaa{2,\cdots,n-1}\shuffle\Spaa{b,a},n|\vec{\pmb{\sigma}}_{n+2})\,,
\eea
then \eref{rela} follows from the general BCJ relation
\bea
\Big(k_a\cdot X_a+k_b\cdot X_b\Big)\,{\cal A}_{\rm BAS}
(1,\Spaa{2,\cdots,n-1}\shuffle\Spaa{b,a},n|\vec{\pmb{\sigma}}_{n+2})=0\,.
\eea
%

%%%%%%%%%%%%%%%%%%%%%%%%%%%%%%%%%%%%%%%
\subsection{Double soft theorem}
\label{subsec-doublesoft}
%%%%%%%%%%%%%%%%%%%%%%%%%%%%%%%%%%%%%

Using the expansion of $(n+2)$-point ${\rm NLSM}\oplus{\rm BAS}$ amplitude with $2$ external pions, provided in \eref{n+2p-expan}, now we can derive the double soft behavior for $k_a\to\tau k_a$, $k_b\to\tau k_b$, $\tau\to0$, i.e., two pions $a$ and $b$ are taken to be soft simultaneously. We are only interested in the subleading soft behavior which is at the $\tau^1$ order. Furthermore, we restrict ourselves to the special orderings $\vec{\pmb\sigma}_{n+2}$ in which $a$ and $b$ are adjacent, therefore $\delta_{ab}\neq0$. As discussed in section \ref{sec-preparations}, for such orderings, we can regard two legs $a$ and $b$ as an individual element encoded as $(ab)$, and introduce $\delta_{i(ab)}$ to denote whether the leg $i$ is adjacent to $(ab)$, namely, $\delta_{i(ab)}=1$ for orderings
$\langle\cdots,i,a,b,\cdots\rangle$ or $\langle\cdots,i,b,a,\cdots\rangle$, $\delta_{i(ab)}=-1$ for orderings $\langle\cdots,a,b,i,\cdots\rangle$ or $\langle\cdots,b,a,i,\cdots\rangle$, and $\delta_{i(ab)}=0$ otherwise.

\begin{figure}
  \centering
  % Requires \usepackage{graphicx}
  \includegraphics[width=16cm]{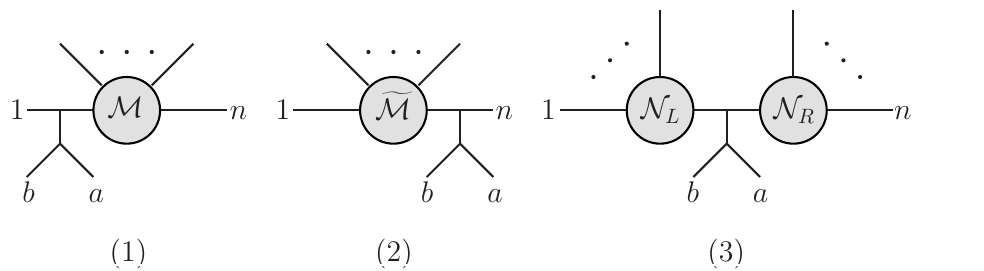} \\
  \caption{Graphs contribute to double soft behavior.}\label{Yes}
\end{figure}

\begin{figure}
  \centering
  % Requires \usepackage{graphicx}
  \includegraphics[width=14cm]{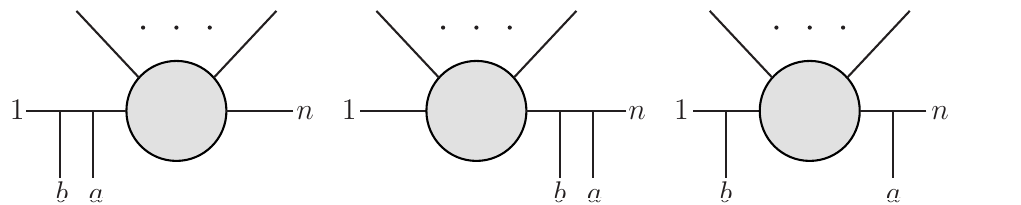} \\
  \caption{First sector of graphs do not contribute to double soft behavior.}\label{No1}
\end{figure}

\begin{figure}
  \centering
  % Requires \usepackage{graphicx}
  \includegraphics[width=16cm]{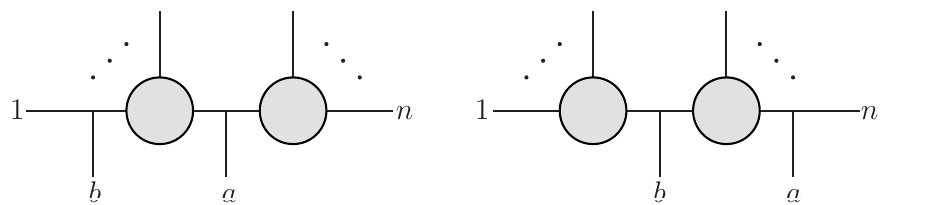} \\
  \caption{Second sector of graphs do not contribute to double soft behavior.}\label{No2}
\end{figure}

It is convenient to fix the legs adjacent to $(ab)$ at the beginning, and figure out the double soft behavior for such special configuration. Since the adjacency can be described by values of symbols such as $\delta_{i(ab)}$, $\delta_{ia}$, $\delta_{ib}$, it is straightforward to generalize the resulted double soft behavior to more general cases by inserting these symbols. Thus, we first consider the ordered sets $\vec{\pmb\sigma}^{(nab1)}_{n+2}=\Spaa{1,\vec{\pmb\sigma}_{n-2},n,a,b}$, where $\vec{\pmb\sigma}_{n-2}$ denotes the ordered set for $(n-2)$ elements in $\{2,\cdots,n-1\}$. For such special orderings $\vec{\pmb\sigma}^{(nab1)}_{n+2}$, Feynman diagrams of BAS amplitudes which contribute to the soft behavior at the $\tau^1$ order are those in Figure.\ref{Yes}. Notice that diagrams in Figure.\ref{No1}
and Figure.\ref{No2} do not contribute, since they are incompatible with the left orderings of BAS amplitudes in the expansion \eref{n+2p-expan}.


The diagram (1) in Figure.\ref{Yes} corresponds to ${\cal A}_{\rm BAS}(1,a,b,2,\cdots,n|\vec{\pmb\sigma}^{(nab1)}_{n+2})$ in the expansion \eref{n+2p-expan}, with the associated coefficient $k_b\cdot P_a\cdot X_a=(k_b\cdot k_a)(k_a\cdot k_1)$. The contribution from this part is given by
\bea
{\cal P}_1&=&-{(\tau^2\,k_b\cdot k_a)\,(\tau\,k_a\cdot k_1)\over (2\tau^2\,k_a\cdot k_b)\,(2\tau\,k_{ab}\cdot k_1+2\tau^2\,k_a\cdot k_b)}\,{\cal M}(\tau)\,,~~\label{P1}
\eea
where the block ${\cal M}$ is labeled in diagram (1) of Figure.\ref{Yes}. In \eref{P1}, the numerator comes from the coefficient $k_b\cdot P_a\cdot X_a$, while the denominator arises from propagators $1/s_{ab}$ and $1/s_{ab1}$. The overall sign $-$ is determined by comparing two orderings carried by the BAS amplitude ${\cal A}_{\rm BAS}(1,a,b,2,\cdots,n|1,\cdots,n,a,b)$. The remaining potential $\pm$ created by turning $\Spaa{2,\cdots,n-1}$ to $\vec{\pmb\sigma}_{n-1}$ is absorbed into ${\cal M}(\tau)$, since it is contained in the $n$-point BAS amplitude
${\cal A}_{\rm BAS}(1,\cdots,n|1,\vec{\pmb\sigma}_{n-1},n)$.
Then, the leading order of ${\cal M}(\tau)$ is obviously ${\cal M}(0)={\cal A}_{\rm BAS}(1,\cdots,n|1,\vec{\pmb\sigma}_{n-1},n)$. Using \eref{P1}, the subleading term of ${\cal P}_1$ is found to be
\bea
{\cal P}^{(1)_{ab}}_1&=&\tau\,{(k_a\cdot k_1)\,(k_a\cdot k_b)\over 4\,(k_{ab}\cdot k_1)^2}\,{\cal A}_{\rm BAS}(1,\cdots,n|1,\vec{\pmb\sigma}_{n-1},n)\nn
& &-\tau\,{k_a\cdot k_1\over 4\,k_{ab}\cdot k_1}\,\Big(k_{ab}\cdot{\partial\over\partial k_1}\,{\cal A}_{\rm BAS}(1,\cdots,n|1,\vec{\pmb\sigma}_{n-1},n)\Big)\,.~~\label{sP1}
\eea
The first line at the r.h.s of \eref{sP1} is obtained by expanding the denominator in \eref{P1}. The second line uses
\bea
\tau\,\Big({\partial\over\partial\tau}\,{\cal M}(\tau)\Big)_{\tau=0}=\tau\,\Big(k_{ab}\cdot{\partial\over\partial(\tau k_{ab})}\,{\cal M}(\tau)\Big)_{\tau=0}=\tau\,\Big(k_{ab}\cdot{\partial\over\partial k_1}\,{\cal M}(\tau)\Big)_{\tau=0}
=\tau\,k_{ab}\cdot{\partial\over\partial k_1}\,{\cal M}(0)\,,
\eea
due to the observation that the parameter $\tau$ enters ${\cal M}(\tau)$ only through the combinatorial momentum $k_1+\tau k_{ab}$,
as can be seen from the corresponding Feynman diagram (1) in Figure.\ref{Yes}.

The situation for diagram (2) in Figure.\ref{Yes} is similar. It corresponds to ${\cal A}_{\rm BAS}(1,\cdots,n-1,a,b,n|\vec{\pmb\sigma}_{n+2}^{nab1})$ in the expansion \eref{n+2p-expan}, and the associated coefficient is $k_b\cdot P_a\cdot X_a=-(k_b\cdot k_a)(k_a\cdot k_{bn})$. The contribution from this diagram is
\bea
{\cal P}_2={(\tau^2\,k_b\cdot k_a)\,(\tau\,k_a\cdot k_n+\tau^2\,k_a\cdot k_b)\over (2\tau^2\,k_a\cdot k_b)\,(2\tau\,k_{ab}\cdot k_n+2\tau^2\,k_a\cdot k_b)}\,\W{\cal M}(\tau)\,,
\eea
then the similar procedure leads to the subleading term
\bea
{\cal P}^{(1)_{ab}}_2&=&\tau\,{(k_b\cdot k_n)\,(k_b\cdot k_a)\over 4\,(k_{ab}\cdot k_n)^2}\,{\cal A}_{\rm BAS}(1,\cdots,n|1,\vec{\pmb\sigma}_{n-2},n)\nn
& &+\tau\,{k_a\cdot k_n\over 4\,k_{ab}\cdot k_n}\,\Big(k_{ab}\cdot{\partial\over\partial k_n}\,{\cal A}_{\rm BAS}(1,\cdots,n|1,\vec{\pmb\sigma}_{n-2},n)\Big)\,.~~\label{sP2}
\eea

The diagram (3) in Figure.\ref{Yes} corresponds to ${\cal A}_{\rm BAS}(1,\cdots,i,a,b,i+1,\cdots,n|\vec{\pmb\sigma}_{n+2}^{(nab1)})$ in the expansion \eref{n+2p-expan} for $i\in\{2,\cdots,n-2\}$, with the coefficients $k_b\cdot P_a\cdot X_a=(k_b\cdot k_a)(k_a\cdot X_a)$. For any fixed $i$, the contribution from this diagram reads
\bea
{\cal P}_{3;i}=-{(\tau^2\,k_b\cdot k_a)\,(\tau\,k_a\cdot X_a)\over 2\tau^2\,k_b\cdot k_a}\,{1\over s_{1\cdots i}}\,{1\over s_{1\cdots i}+2\tau\,k_{ab}\cdot k_{1\cdots i}}\,{\cal N}_L\,{\cal N}_R(\tau)\,,
\eea
or equivalently
\bea
\W {\cal P}_{3;i}=-{(\tau^2\,k_b\cdot k_a)\,(\tau\,k_a\cdot X_a)\over 2\tau^2\,k_b\cdot k_a}\,{1\over s_{(i+1)\cdots n}+2\tau\,k_{ab}\cdot k_{(i+1)\cdots n}}\,{1\over s_{(i+1)\cdots n}}\,{\cal N}_L(\tau)\,{\cal N}_R\,,
\eea
where two blocks ${\cal N}_L$ and ${\cal N}_R$ are labeled in diagram (3) in Figure.\ref{Yes}. Two expressions ${\cal P}_{3;i}$ and $\W {\cal P}_{3;i}$ are related by momentum conservation.
At the desired $\tau^1$ order, we find
\bea
{\cal P}^{(0)_{ab}}_{3;i}&=&{\tau\over4}\,\Big(k_a\cdot {\partial\over\partial k_1}\,{1\over s_{1\cdots i}}\Big)\,{\cal N}_L\,{\cal N}_R(0)\,,
\eea
and
\bea
\W{\cal P}^{(0)_{ab}}_{3;i}&=&-{\tau\over4}\,\Big(k_a\cdot {\partial\over\partial k_n}\,{1\over s_{(i+1)\cdots n}}\Big)\,{\cal N}_L(0)\,{\cal N}_R\,.
\eea
Here a subtle point is that although $k_1^2$ vanishes due to the on-shell condition, it contributes to $\partial_{k^\mu_1}s_{1\cdots i}$ since $\partial_{k_1^\mu}k_1^2=2k_{1\mu}$. The situation for
$\partial_{k^\mu_n}\, s_{(i+1)\cdots n}$ is analogous. Now we can sum over all $i\in\{2,\cdots,n-2\}$ to get
\bea
{\cal P}^{(0)_{ab}}_{3}&=&\sum_{i=2}^{n-2}\,\big({\cal P}^{(0)_{ab}}_{3;i}+\W{\cal P}^{(0)_{ab}}_{3;i}\big)\nn
&=&{\tau\over4}\,\Big(k_a\cdot{\partial\over\partial k_1}-k_a\cdot{\partial\over\partial k_n}\Big)\,{\cal A}_{\rm BAS}(1,\cdots,n|1,\vec{\pmb\sigma}_{n-2},n)\,.~~\label{sP3}
\eea
Notice that ${\cal P}^{(0)_{ab}}_{3;i}$ and $\W{\cal P}^{(0)_{ab}}_{3;i}$ need to be summed together. The reason is, one can always alter the expression of propagators by using momentum conservation. For any particular expression, one of ${\cal P}^{(0)_{ab}}_{3;i}$ and $\W{\cal P}^{(0)_{ab}}_{3;i}$ vanishes, but the summation of them always gives rise to the correct answer. Thus, the result which is compatible with momentum conservation can only be achieved by summing them together.

Combining ${\cal P}^{(0)_{ab}}_{1}$, ${\cal P}^{(0)_{ab}}_{2}$ and ${\cal P}^{(0)_{ab}}_{3}$ in \eref{sP1}, \eref{sP2} and \eref{sP3} together, we arrive at
\bea
& &{\cal A}^{(1)_{ab}}_{{\rm N}\oplus{\rm B}}(1,\cdots,n;\{a,b\}|\vec{\pmb\sigma}_{n+2}^{nab1})\nn
&=&-{\tau\over4\,k_{ab}\cdot k_1}\,\Big(k_1\cdot L_{ab}\cdot{\partial\over\partial k_1}-k_n\cdot L_{ab}\cdot{\partial\over\partial k_n}\Big)\,{\cal A}_{\rm BAS}(1,\cdots,n|1,\vec{\pmb\sigma}_{n-2},n)\nn
& &+\tau\,\Big({(k_a\cdot k_1)\,(k_a\cdot k_b)\over 4\,(k_{ab}\cdot k_1)^2}+{(k_b\cdot k_n)\,(k_a\cdot k_b)\over 4\,(k_{ab}\cdot k_n)^2}\Big)\,{\cal A}_{\rm BAS}(1,\cdots,n|1,\vec{\pmb\sigma}_{n-2},n)\,,
\eea
where $L^{\mu\nu}_a\equiv k^\mu_a k^\nu_b-k^\mu_b k^\nu_a$.
The above double soft behavior holds for particular orderings $\vec{\pmb\sigma}_{n+2}^{nab1}=\{1,\vec{\pmb\sigma}_{n-2},n,a,b\}$. As discussed previously, we can insert symbols $\delta_{ij}$ and $\delta_{i(ab)}$ to get the more general formula
\bea
& &{\cal A}^{(1)_{ab}}_{{\rm N}\oplus{\rm B}}(1,\cdots,n;\{a,b\}|\vec{\pmb\sigma}_{n+2}^{ab})\nn
&=&\Big(S^{(1)_{ab}}_D+S^{(1)_{ab}}_C\Big)\,{\cal A}_{\rm BAS}(1,\cdots,n|\vec{\pmb\sigma}_{n+2}^{ab}\setminus\{a,b\})\,,~~\label{doublesoft-theo}
\eea
where
\bea
S^{(1)_{ab}}_D&=&\tau\,\delta_{ab}\,\Big(\sum_{i=1}^n\,{\delta_{i(ab)}\over4\,k_{ab}\cdot k_i}\,k_i\cdot L_{ab}\cdot{\partial\over\partial k_i}\Big)\,,\nn
S^{(1)_{ab}}_C&=&\tau\,\Big(\sum_{i=1}^n\,{\delta_{ab}\,\delta_{bi}\,(k_i\cdot k_a)\,(k_a\cdot k_b)\over 4\,(k_{ab}\cdot k_i)^2}+\sum_{i=1}^n\,{\delta_{ba}\,\delta_{ai}\,(k_i\cdot k_b)\,(k_b\cdot k_a)\over 4\,(k_{ab}\cdot k_i)^2}\Big)\,.~~\label{doublesoft-fac}
\eea
The double soft behavior in \eref{doublesoft-theo} holds for orderings $\vec{\pmb\sigma}_{n+2}^{ab}$ in which pions $a$ and $b$ are adjacent.
The soft factor is decomposed into two parts $S^{(1)_{ab}}_D$ and $S^{(1)_{ab}}_C$ as in \eref{doublesoft-fac}, where $S^{(1)_{ab}}_D$ is a differential operator while $S^{(1)_{ab}}_C$ is not.

Before ending this subsection, we display the effect of acting operator $S^{(1)_{ab}}_D$ on the Lorentz invariant $k_j\cdot k_g$, which will be used frequently in subsequent sections. Using the definition of $S^{(1)_{ab}}_D$ in \eref{doublesoft-fac}, we have
\bea
S^{(1)_{ab}}_D\,(k_j\cdot k_g)&=&\tau\,\delta_{ab}\,\Big({\delta_{j(ab)}\over 4\,k_{ab}\cdot k_j}\,k_j\cdot L_{ab}\cdot k_g+
{\delta_{g(ab)}\over 4\,k_{ab}\cdot k_g}\,k_g\cdot L_{ab}\cdot k_j\Big)\nn
&=&\tau\,\delta_{ab}\,\Big({\delta_{j(ab)}\over 4\,k_{ab}\cdot k_j}-
{\delta_{g(ab)}\over 4\,k_{ab}\cdot k_g}\Big)\,k_j\cdot L_{ab}\cdot k_g\nn
&=&\tau\,\delta_{ab}\,\Big({\delta_{j(ab)}\over 4\,k_{ab}\cdot k_j}-
{\delta_{g(ab)}\over 4\,k_{ab}\cdot k_g}\Big)\,{k_j\cdot P_a\cdot P_b\cdot k_g-k_j\cdot P_b\cdot P_a\cdot k_g\over k_a\cdot k_b}\nn
&=&\tau\,{\delta_{ab}\over 2k_a\cdot k_b}\,\Big({\delta_{g(ab)}\over 2\,k_{ab}\cdot k_g}+
{\delta_{(ab)j}\over 2\,k_{ab}\cdot k_j}\Big)\,k_j\cdot P_b\cdot P_a\cdot k_g\nn
& &+\tau\,{\delta_{ba}\over 2k_a\cdot k_b}\,\Big({\delta_{g(ab)}\over 2\,k_{ab}\cdot k_g}+
{\delta_{(ab)j}\over 2\,k_{ab}\cdot k_j}\Big)\,k_j\cdot P_a\cdot P_b\cdot k_g\,.~~\label{action-doublesoft}
\eea
It seems that the final formula in \eref{action-doublesoft} is more complicated than the first line. However, this formula manifests the connection between $S^{(1)_{ab}}_D\,(k_j\cdot k_g)$ and the leading double soft behavior of BAS amplitudes in \eref{soft-2s1} and \eref{soft-fac-2s}. The advantage of such connection will be seen in section \ref{sec-constructNLSM} and section \ref{sec-higherderi}.

%%%%%%%%%%%%%%%%%%%%%%%%%%%%%%%%%%%%%%%%%%%%%
\section{Construct NLSM$\oplus$BAS and NLSM amplitudes by inverting double soft theorem}
\label{sec-constructNLSM}
%%%%%%%%%%%%%%%%%%%%%%%%%%%%%%%%%%%%%%%%%%%%%%

In this section, we invert the subleading double soft theorem given in \eref{doublesoft-theo} and \eref{doublesoft-fac}, to construct general ${\rm NLSM}\oplus{\rm BAS}$ amplitudes and NLSM amplitudes. This method is based on the observation that the $4$-point NLSM amplitudes can be reinterpreted as ${\rm NLSM}\oplus{\rm BAS}$ amplitudes. In the soft limit $\tau\to0$ with $k_a\to\tau k_a$, $k_b\to\tau k_b$, each propagator $1/s_{abc}$ becomes on-shell, thus the quadrivalent pion-pion-pion-pion and pion-pion-BAS-BAS vertices are reduced to on-shell amplitudes, then the equivalence between $4$-point NLSM and ${\rm NLSM}\oplus{\rm BAS}$ amplitudes indicates the equivalent soft behaviors for soft
pions attached to corresponding vertices. Therefore, it is natural to assume that the double soft behavior given in \eref{doublesoft-theo} and \eref{doublesoft-fac} is universal, valid for any NLSM or ${\rm NLSM}\oplus{\rm BAS}$ amplitude.

To explain the reason for choosing subleading order rather than the leading one, let us take the NLSM amplitudes as example. Near $s_{abc}=0$, the full $n$-point amplitude is factorized into a $4$-point on-shell NLSM amplitude and a $(n-2)$-point one. Using the unique solution for $4$-point NLSM amplitudes in \eref{4p1} and \eref{4p2}, we find the soft behavior of each $4$-point amplitude contains two parts with different leading orders, one is the $\tau^1$ order, another one is the $\tau^2$. For instance, consider ${\cal A}_{\rm NLSM}(1,2,3,4)=s+t$, suppose we take pions $1$ and $2$ to be soft, then the leading order of $s$ is the $\tau^2$ order, while the leading order of $t$ is the $\tau^1$. Together with the propagator $1/s_{abc}$ which contributes $\tau^{-1}$, we see the two parts of $4$-point NLSM amplitude correspond to terms with leading orders $\tau^0$ and $\tau^1$ respectively. According to the above estimation, we should consider the double soft behavior at the $\tau^1$ order, to detect both of them. The situation of ${\rm NLSM}\oplus{\rm BAS}$ amplitudes is similar.

%%%%%%%%%%%%%%%%%%%%%%%%%%%%%%%%%
\subsection{General ${\rm NLSM}\oplus{\rm BAS}$ amplitudes}
\label{subsec-NB}
%%%%%%%%%%%%%%%%%%%%%%%%%%%%%%%%%

Let us consider the $(n+4)$-point ${\rm NLSM}\oplus{\rm BAS}$ amplitude ${\cal A}_{{\rm N}\oplus{\rm B}}(1,\cdots,n;\{a,b,c,d\}|\vec{\pmb\sigma}_{n+4})$ with $n$ external BAS scalars in $\{1,\cdots,n\}$, and $4$ external pions encoded as $a$, $b$, $c$ and $d$.
We take the double soft limit $k_c\to\tau k_c$, $k_d\to\tau k_d$, $\tau\to 0$. The double soft theorem in \eref{doublesoft-theo} forces the following subleading soft behavior
\bea
& &{\cal A}^{(1)_{cd}}_{{\rm N}\oplus{\rm B}}(1,\cdots,n;\{a,b,c,d\}|\vec{\pmb\sigma}_{n+4})\nn
&=&\big(S^{(1)_{cd}}_D+S^{(1)_{cd}}_C\big)\,{\cal A}_{{\rm N}\oplus{\rm B}}(1,\cdots,n;\{a,b\}|\vec{\pmb\sigma}_{n+4}\setminus\{c,d\})\nn
&=&\big(S^{(1)_{cd}}_D+S^{(1)_{cd}}_C\big)\,\Big((k_b\cdot P_a\cdot X_a)\,{\cal A}_{\rm BAS}(1,\Spaa{a,b}\shuffle\Spaa{2,\cdots,n-1},n|\vec{\pmb\sigma}_{n+4}\setminus\{c,d\})\Big)\nn
&=&(k_b\cdot P_a\cdot X_a)\,\Big(\big(S^{(1)_{cd}}_D+S^{(1)_{cd}}_C\big)\,{\cal A}_{\rm BAS}(1,\Spaa{a,b}\shuffle\Spaa{2,\cdots,n-1},n|\vec{\pmb\sigma}_{n+4}\setminus\{c,d\})\Big)\nn
& &+\Big(S^{(1)_{cd}}_D\,(k_b\cdot P_a\cdot X_a)\Big)\,{\cal A}_{\rm BAS}(1,\Spaa{a,b}\shuffle\Spaa{2,\cdots,n-1},n|\vec{\pmb\sigma}_{n+4}\setminus\{c,d\})\,.~~\label{n+4p-soft}
\eea
In \eref{n+4p-soft}, the second equality is obtained by substituting the expansion of $(n+2)$-point ${\rm NLSM}\oplus{\rm BAS}$ amplitude in \eref{n+2p-expan}, the third uses the observation that $S^{(1)_{cd}}_D$ is a differential operator which satisfies Leibnitz rule. One can figure out each part in \eref{n+4p-soft} as
\bea
D_1&=&(k_b\cdot P_a\cdot X_a)\,\Big(\big(S^{(1)_{cd}}_D+S^{(1)_{cd}}_C\big)\,{\cal A}_{\rm BAS}(1,\Spaa{a,b}\shuffle\Spaa{2,\cdots,n-1},n|\vec{\pmb\sigma}_{n+4}\setminus\{c,d\})\Big)\nn
&=&(k_b\cdot P_a\cdot X_a)\,{\cal A}^{(1)_{cd}}_{{\rm N}\oplus{\rm B}}(1,\Spaa{a,b}\shuffle\Spaa{2,\cdots,n-1},n;\{c,d\}|\vec{\pmb\sigma}_{n+4})\,,~~\label{B1}
\eea
based on inverting the double soft theorem, and
\bea
D_2&=&\Big(S^{(1)_{cd}}_D\,(k_b\cdot P_a\cdot X_a)\Big)\,{\cal A}_{\rm BAS}(1,\Spaa{a,b}\shuffle\Spaa{2,\cdots,n-1},n|\vec{\pmb\sigma}_{n+4}\setminus\{c,d\})\nn
&=&\tau^4\,(k_b\cdot P_a\cdot P_d\cdot P_c\cdot X_c)\,{\cal A}^{(0)_{cd}}_{\rm BAS}(1,\Spaa{c,d,a,b}\shuffle\Spaa{2,\cdots,n-1},n|\vec{\pmb\sigma}_{n+4})\nn
& &+\tau^4\,(k_b\cdot P_a\cdot P_c\cdot P_d\cdot X_d)\,{\cal A}^{(0)_{cd}}_{\rm BAS}(1,\Spaa{d,c,a,b}\shuffle\Spaa{2,\cdots,n-1},n|\vec{\pmb\sigma}_{n+4})\nn
& &+\tau^4\,(k_b\cdot P_d\cdot P_c\cdot P_a\cdot X_a)\,{\cal A}^{(0)_{cd}}_{\rm BAS}(1,\Spaa{a,c,d,b}\shuffle\Spaa{2,\cdots,n-1},n|\vec{\pmb\sigma}_{n+4})\nn
& &+\tau^4\,(k_b\cdot P_c\cdot P_d\cdot P_a\cdot X_a)\,{\cal A}^{(0)_{cd}}_{\rm BAS}(1,\Spaa{a,d,c,b}\shuffle\Spaa{2,\cdots,n-1},n|\vec{\pmb\sigma}_{n+4})\,,~~\label{B2}
\eea
by using the result in \eref{action-doublesoft}, the double soft theorem for BAS amplitudes in \eref{soft-2s1} and \eref{soft-fac-2s}, as well as the technic in \eref{tech-double}. Substituting \eref{B1} and \eref{B2} into \eref{n+4p-soft}, we obtain
\bea
& &{\cal A}^{(1)_{cd}}_{{\rm N}\oplus{\rm B}}(1,\cdots,n;\{a,b,c,d\}|\vec{\pmb\sigma}_{n+4})\nn
&=&(k_b\cdot P_a\cdot X_a)\,{\cal A}^{(1)_{cd}}_{{\rm N}\oplus{\rm B}}(1,\Spaa{a,b}\shuffle\Spaa{2,\cdots,n-1},n;\{c,d\}|\vec{\pmb\sigma}_{n+4})\nn
& &+\tau^4\,(k_b\cdot P_a\cdot P_d\cdot P_c\cdot X_c)\,{\cal A}^{(0)_{cd}}_{\rm BAS}(1,\Spaa{c,d,a,b}\shuffle\Spaa{2,\cdots,n-1},n|\vec{\pmb\sigma}_{n+4})\nn
& &+\tau^4\,(k_b\cdot P_a\cdot P_c\cdot P_d\cdot X_d)\,{\cal A}^{(0)_{cd}}_{\rm BAS}(1,\Spaa{d,c,a,b}\shuffle\Spaa{2,\cdots,n-1},n|\vec{\pmb\sigma}_{n+4})\nn
& &+\tau^4\,(k_b\cdot P_d\cdot P_c\cdot P_a\cdot X_a)\,{\cal A}^{(0)_{cd}}_{\rm BAS}(1,\Spaa{a,c,d,b}\shuffle\Spaa{2,\cdots,n-1},n|\vec{\pmb\sigma}_{n+4})\nn
& &+\tau^4\,(k_b\cdot P_c\cdot P_d\cdot P_a\cdot X_a)\,{\cal A}^{(0)_{cd}}_{\rm BAS}(1,\Spaa{a,d,c,b}\shuffle\Spaa{2,\cdots,n-1},n|\vec{\pmb\sigma}_{n+4})\,,
\eea
which implies
\bea
& &{\cal P}^{cd}_{{\rm N}\oplus{\rm B}}(1,\cdots,n;\{a,b,c,d\}|\vec{\pmb\sigma}_{n+4})\nn
&=&(k_b\cdot P_a\cdot X_a)\,{\cal A}_{{\rm N}\oplus{\rm B}}(1,\Spaa{a,b}\shuffle\Spaa{2,\cdots,n-1},n;\{c,d\}|\vec{\pmb\sigma}_{n+4})\nn
& &+(k_b\cdot P_a\cdot P_d\cdot P_c\cdot X_c)\,{\cal A}_{\rm BAS}(1,\Spaa{c,d,a,b}\shuffle\Spaa{2,\cdots,n-1},n|\vec{\pmb\sigma}_{n+4})\nn
& &+(k_b\cdot P_a\cdot P_c\cdot P_d\cdot X_d)\,{\cal A}_{\rm BAS}(1,\Spaa{d,c,a,b}\shuffle\Spaa{2,\cdots,n-1},n|\vec{\pmb\sigma}_{n+4})\nn
& &+(k_b\cdot P_d\cdot P_c\cdot P_a\cdot X_a)\,{\cal A}_{\rm BAS}(1,\Spaa{a,c,d,b}\shuffle\Spaa{2,\cdots,n-1},n|\vec{\pmb\sigma}_{n+4})\nn
& &+(k_b\cdot P_c\cdot P_d\cdot P_a\cdot X_a)\,{\cal A}_{\rm BAS}(1,\Spaa{a,d,c,b}\shuffle\Spaa{2,\cdots,n-1},n|\vec{\pmb\sigma}_{n+4})\,,
\eea
where ${\cal P}^{cd}_{{\rm N}\oplus{\rm B}}(1,\cdots,n;\{a,b,c,d\}|\vec{\pmb\sigma}_{n+4})$ serves as the special part of the full amplitude, which can be detected by applying the double soft theorem in \eref{doublesoft-theo} and \eref{doublesoft-fac} for $k_c\to\tau k_c$, $k_d\to\tau k_d$. This double soft theorem is valid for special cases those $c$ and $d$ are adjacent legs in $\vec{\pmb\sigma}_{n+4}$, the soft factor carries $\delta_{cd}$ therefore vanishes for any ordering $\vec{\pmb\sigma}_{n+4}$ with separated $c$ and $d$. Thus, the complete amplitude ${\cal A}_{{\rm N}\oplus{\rm B}}(1,\cdots,n;\{a,b,c,d\}|\vec{\pmb\sigma}_{n+4})$ can be recovered by breaking the adjacency of $c$ and $d$, namely,
\bea
& &{\cal A}_{{\rm N}\oplus{\rm B}}(1,\cdots,n;\{a,b,c,d\}|\vec{\pmb\sigma}_{n+4})\nn
&=&\sum_{\vec{\pmb\a},|\pmb\a|={\rm odd}}\,(k_b\cdot P_{\vec{\pmb\a}}\cdot X_{\vec{\pmb\a}})\,{\cal A}_{{\rm N}\oplus{\rm B}}(1,\Spaa{\vec{\pmb\a},b}\shuffle\Spaa{2,\cdots,n-1},n;\{a,c,d\}\setminus\pmb\a|\vec{\pmb\sigma}_{n+4})\,.~~\label{n+4p-amp}
\eea
In the above expansion \eref{n+4p-amp}, $\pmb\a$ stands for subsets of $\{a,c,d\}$, and each ordered set $\vec{\pmb\a}$ is obtained by giving ordering to elements in the corresponding $\pmb\a$. For any $\vec{\pmb\a}=\Spaa{\a_1,\cdots,\a_k}$, the tensor $P_{\vec{\pmb\a}}$ is defined as
\bea
P^{\mu\nu}_{\vec{\pmb\a}}\equiv(P_{\a_k}\cdot P_{\a_{k-1}}\cdots P_{\a_2}\cdot P_{\a_1})^{\mu\nu}\,.
\eea
The combinatorial momentum $X_{\vec{\pmb\a}}$ is defined as $X_{\a_1}$ where $\a_1$ is the first element in $\vec{\pmb\a}$.
The notation $|\pmb\a|$ stands for the number of elements included in $\pmb\a$, thus the summation is for all inequivalent $\vec{\pmb\a}$ which include the odd numbers of elements. When $\pmb\a=\{a,c,d\}$, the corresponding ${\rm NLSM}\oplus{\rm BAS}$ amplitudes in the expansion are reduced to BAS ones.

Repeating the above construction recursively, one can generalize the expansion in \eref{n+4p-amp} to ${\rm NLSM}\oplus{\rm BAS}$ amplitudes with arbitrary numbers of external BAS scalars and external pions,
\bea
& &{\cal A}_{{\rm N}\oplus{\rm B}}(1,\cdots,n;\pmb{p}_{2m}|\vec{\pmb\sigma}_{n+2m})\nn
&=&\sum_{\vec{\pmb\a},|\pmb\a|={\rm odd}}\,(k_f\cdot P_{\vec{\pmb\a}}\cdot X_{\vec{\pmb\a}})\,{\cal A}_{{\rm N}\oplus{\rm B}}(1,\Spaa{\vec{\pmb\a},f}\shuffle\Spaa{2,\cdots,n-1},n;\pmb{p}_{2m}\setminus\{\pmb\a\cup f\}|\vec{\pmb\sigma}_{n+2m})\,.~~\label{n+mp-amp}
\eea
In the above, $\pmb{p}_{2m}$ denotes the unordered set of $2m$ external pions, and the fiducial pion $b$ in \eref{n+4p-amp} is relabeled as $f$.
Through the derivation paralleled to that from \eref{n+4p-soft} to \eref{n+4p-amp}, it is straightforward to show that the general expansion
\eref{n+mp-amp} holds for amplitudes ${\cal A}_{{\rm N}\oplus{\rm B}}(1,\cdots,n;\pmb{p}_{2(k+1)}|\vec{\pmb\sigma}_{n+2(k+1)})$, if it is satisfied
by amplitudes ${\cal A}_{{\rm N}\oplus{\rm B}}(1,\cdots,n;\pmb{p}_{2k}|\vec{\pmb\sigma}_{n+2k})$.
Formally, the summation in \eref{n+mp-amp} can also be extended to that for all inequivalent ordered sets $\vec{\pmb\a}$, since the effective terms are still those $\pmb\a$ include odd number of pions. The reason is, any ${\rm NLSM}\oplus{\rm BAS}$ amplitude with odd number of external pions  vanishes automatically, as will be explained at the end of this subsection.

The resulted expansion in \eref{n+mp-amp} can be verified by employing the transmutation operator
which turns external gluons to pions \cite{Cheung:2017ems,Zhou:2018wvn,
Bollmann:2018edb}. Such operator is defined as
\bea
{\cal L}_{2m}\equiv\prod_{i\in\pmb{g}_{2m}}\,{\cal L}_i=\bar{{\cal L}}+\cdots\,,~~\label{Defin-L}
\eea
where
\bea
{\cal L}_i&\equiv&\sum_{j\in\pmb{g}_{2m}\setminus i}\,(k_i\cdot k_j)\,\partial_{\epsilon_i\cdot k_j}\,,\nn
\bar{{\cal L}}_{2m}&\equiv&-\sum_{\rho\in{\rm pair}}\,\prod_{i,j\in\rho}\,(k_i\cdot k_j)\,\partial_{\epsilon_i\cdot\epsilon_j}\,,
\eea
and $\pmb{g}_{2m}$ denotes the set of $2m$ gluons. In the definition of $\bar{{\cal L}}_{2m}$, $\rho$ stands for a partition of $\pmb{g}_{2m}$ into pairs without regard to the order, and the summation is for all inequivalent partitions. Two operators ${\cal L}_{2m}$ and $\bar{{\cal L}}_{2m}$ do not equivalent to each other at the algebraic level, but gives the same effect when acting on single-trace Yang-Mills-scalar (YMS) amplitudes with $2m$ external gluons\footnote{The transmutation relation which turns pure Yang-Mills amplitudes to pure NLSM ones is proved in \cite{Zhou:2018wvn,
Bollmann:2018edb} by using Cachazo-He-Yuan (CHY) formula. The proof in \cite{Zhou:2018wvn,
Bollmann:2018edb} can be directly extended to the current case of transmuting YMS amplitudes to ${\rm NLSM}\oplus{\rm BAS}$ ones, by applying totally the same derivation to CHY integrands for YMS and ${\rm NLSM}\oplus{\rm BAS}$ theories.},
\bea
{\cal L}_{2m}\,{\cal A}_{\rm YMS}(1,\cdots,n;\pmb{g}_{2m}|\vec{\pmb\sigma}_{n+2m})=\bar{{\cal L}}_{2m}\,{\cal A}_{\rm YMS}(1,\cdots,n;\pmb{g}_{2m}|\vec{\pmb\sigma}_{n+2m})={\cal A}_{{\rm N}\oplus{\rm B}}(1,\cdots,n;\pmb{p}_{2m}|\vec{\pmb\sigma}_{n+2m})\,.~~\label{trans}
\eea
Then, by using the universal expansion of single-trace YMS amplitudes \cite{Fu:2017uzt}
\bea
& &{\cal A}_{\rm YMS}(1,\cdots,n;\pmb{g}_{2m}|\vec{\pmb\sigma}_{n+2m})\nn
&=&(\epsilon_f\cdot F_{\vec{\pmb\a}}\cdot X_{\vec{\pmb\a}})\,{\cal A}_{\rm YMS}(1,\Spaa{2,\cdots,n-1}\shuffle\vec{\pmb\a} ,n;\pmb{g}_{2m}\setminus\pmb\a|\vec{\pmb\sigma}_{n+2m})\,,~~\label{expan-YMS}
\eea
as well as the transmutation relation in \eref{trans}, the expansion of ${\rm NLSM}\oplus{\rm BAS}$ amplitudes in \eref{n+mp-amp} is reproduced.
In \eref{expan-YMS}, the tensor $F_{\vec{\pmb\a}}$ is defined as
\bea
F^{\mu\nu}_{\vec{\pmb\a}}\equiv (f_{\a_k}\cdot f_{\a_{k-1}}\cdots f_{\a_2}\cdot f_{\a_1})^{\mu\nu}\,,~~~~{\rm for}~\vec{\pmb\a}=\Spaa{\a_1,\cdots,\a_k}\,,
\eea
where $f_i^{\mu\nu}\equiv k_i^\mu\epsilon_i^\nu-\epsilon_i^\mu k_i^\nu$.

Substituting the expansion in \eref{n+mp-amp} into itself iteratively, one can obtain another elegant formula for ${\rm NLSM}\oplus{\rm BAS}$ amplitudes,
\bea
{\cal A}_{{\rm N}\oplus{\rm B}}(1,\cdots,n;\pmb{p}_{2m}|\vec{\pmb\sigma}_{n+2m})
&=&\sum_{\vec{\pmb{p}}_{2m}}\,\Big(\prod_{i=1}^{2m}\,(k_{p_i}\cdot X_{k_{p_i}})\Big)\,{\cal A}_{\rm BAS}(1,\vec{\pmb{p}}_{2m}\shuffle\{2,\cdots,n-1\},n|\vec{\pmb\sigma}_{n+2m})\,,~~\label{NB-BAS}
\eea
which expands ${\cal A}_{{\rm N}\oplus{\rm B}}(1,\cdots,n;\pmb{p}_{2m}|\vec{\pmb\sigma}_{n+2m})$ to pure BAS amplitudes.
In \eref{NB-BAS}, pions in the set $\pmb{p}_{2m}$ are encoded as $\pmb{p}_{2m}=\{p_1,\cdots,p_{2m}\}$, and the ordered sets $\vec{\pmb{p}}_{2m}$ are generated from
$\pmb{p}_{2m}$ by giving orderings. Suppose we turn $2m$ to $(2m-1)$, then the formula in \eref{NB-BAS} vanishes, as proved in \cite{Du:2018khm}. Thus we conclude that the effective  ${\rm NLSM}\oplus{\rm BAS}$ amplitudes are those with even number of external pions. The expansion in \eref{NB-BAS} manifests the permutation invariance among external pions, which is not explicit in \eref{n+mp-amp} since a fiducial pion $f$ is introduced.

%%%%%%%%%%%%%%%%%%%%%%%%%%%%%%%%%
\subsection{General NLSM amplitudes}
\label{subsec-npNLSM}
%%%%%%%%%%%%%%%%%%%%%%%%%%%%%%%%%

In section \ref{subsec-4pNLSM}, we argued that the $4$-point NLSM amplitude with fixed ordering $\vec{\pmb\sigma}_4$ is equivalent to the ${\rm NLSM}\oplus{\rm BAS}$ amplitude with two external BAS scalars, two external pions, and the same ordering $\vec{\pmb\sigma}_4$. Thus, suppose we use the recursive technic in the previous subsection \ref{subsec-NB} to construct the $6$-point NLSM amplitude ${\cal A}_{\rm NLSM}(\vec{\pmb\sigma}_6)$ from the $4$-point one ${\cal A}_{\rm NLSM}(\vec{\pmb\sigma}_4)$, the process is exactly the same as constructing the $6$-point ${\rm NLSM}\oplus{\rm BAS}$ amplitude ${\cal A}_{{\rm N}\oplus{\rm B}}(1,2;\{3,4,5,6\})$ from the $4$-point one ${\cal A}_{{\rm N}\oplus{\rm B}}(1,2;\{3,4\})$. Applying the above argument to higher-point cases recursively, one can conclude that the $2m$-point NLSM amplitudes satisfy
\bea
{\cal A}_{\rm NLSM}(\vec{\pmb\sigma}_{2m})={\cal A}_{{\rm N}\oplus{\rm B}}(1,2m;\{2,\cdots,2m-1\}|\vec{\pmb\sigma}_{2m})\,,
\eea
therefore
\bea
{\cal A}_{\rm NLSM}(\vec{\pmb\sigma}_{2m})=\sum_{\vec{\pmb\a}}\,(k_f\cdot P_{\vec{\pmb\a}}\cdot k_1)\,{\cal A}_{{\rm N}\oplus{\rm B}}(1,\vec{\pmb\a},f,2m;\{2,\cdots,2m-1\}\setminus\{\pmb\a\cup f\}|\vec{\pmb\sigma}_{2m})\,,~~\label{expan-NLSM-recur}
\eea
where the observation $X_{\vec{\pmb\a}}=k_1$ is used.
Meanwhile, the expansion in \eref{NB-BAS} is now turned to
\bea
{\cal A}_{\rm NLSM}(\vec{\pmb\sigma}_{2m})=\sum_{\vec{\pmb\sigma}_{2m-2}}\,\Big(\prod_{i=2}^{2m-1}\,(k_i\cdot X_i)\Big)\,{\cal A}_{\rm BAS}(1,\vec{\pmb\sigma}_{2m-2},2m|\vec{\pmb\sigma}_{2m})\,,~~\label{expan-NLSM-BAS}
\eea
where the summation is for all inequivalent orderings $\vec{\pmb\sigma}_{2m-2}$ for $(2m-2)$ elements in $\{2,\cdots,2m-1\}$.
The expansion in \eref{expan-NLSM-BAS} was also found in \cite{Feng:2019tvb,Zhou:2023quv,Zhou:2023vzl}, by applying different approaches.

As explained in \cite{Feng:2019tvb}, the expansion of NLSM amplitudes in \eref{expan-NLSM-BAS} can be directly generated from the expansion of YM amplitudes,
by applying the transmutation operator,
\bea
{\cal L}_{2m-2}\,\partial_{\epsilon_1\cdot\epsilon_{2m}}\,{\cal A}_{\rm YM}(\vec{\pmb\sigma}_{2m})={\cal A}_{\rm NLSM}(\vec{\pmb\sigma}_{2m})\,,~~\label{YM0-NLSM0}
\eea
where ${\cal L}_{2m-2}$ is the transmutation operator given in \eref{Defin-L}, this time is defined for $(2m-2)$ gluons in the set $\pmb{g}_{2m}\setminus\{1,2m\}$.
In \cite{Zhou:2018wvn,Bollmann:2018edb}, the above transmutation is proved by employing CHY formula.

%%%%%%%%%%%%%%%%%%%%%%%%%%%%%%%%%%%%
\section{Pion amplitudes with leading higher-derivative correction}
\label{sec-higherderi}
%%%%%%%%%%%%%%%%%%%%%%%%%%%%%%%%%%%%

In this section, we construct the ordered pion amplitudes which contain the single insertion of local higher-derivative operator, with mass dimension ${\cal D}+4$, where ${\cal D}$ encodes the mass dimension
of ordinary NLSM amplitudes.
In subsection \ref{subsec-4p-higher}, we bootstrap the $4$-point amplitudes with mass dimension ${\cal D}+4$, and discuss the
double soft behavior of general pion amplitudes with mass dimension ${\cal D}+4$, which serves as the foundation of the construction in subsection \ref{subsec-higher-point}. Then, by inverting the double soft theorem in \eref{doublesoft-theo} and \eref{doublesoft-fac}, we construct the higher-point amplitudes with mass dimension ${\cal D}+4$ in subsection \ref{subsec-higher-point}.

%%%%%%%%%%%%%%%%%%%%%%%%%%%%%%
\subsection{$4$-point NLSM$^{+4}$ amplitudes and double soft behavior}
\label{subsec-4p-higher}
%%%%%%%%%%%%%%%%%%%%%%%%%%%%%%%%%

Our previous construction for ${\rm NLSM}\oplus{\rm BAS}$ and NLSM amplitudes starts from bootstrapping $4$-point amplitudes, based on general constraints such as mass dimension and symmetry among different channels. It is natural to ask, which consistent $4$-point scalar amplitudes without any pole can be found if the mass dimension is modified.

Let us restrict the consideration to the ordered amplitudes which satisfy both KK and BCJ relations. The $4$-point NLSM amplitudes have mass dimension $2$. For mass dimension $4$, there is no consistent solution of $4$-point amplitudes. For mass dimension $6$, we find the unique solution
\bea
{\cal A}_{\rm NLSM^{+4}}(1,2,3,4)&=&s^3+2s^2t+2st^2+t^3\,,\nn
{\cal A}_{\rm NLSM^{+4}}(1,3,2,4)&=&t^3+2t^2u+2tu^2+u^3\,,\nn
{\cal A}_{\rm NLSM^{+4}}(1,2,4,3)&=&s^3+2s^2u+2su^2+u^3\,,~~\label{solu-4p-N+4}
\eea
which simultaneously satisfy KK and BCJ relations. Here we used the subscript ${\rm NLSM}^{+4}$ to denote amplitudes for massless pions with mass dimension ${\cal D}+4$. Using the solution \eref{solu-4p-N+4} and the $4$-point NLSM amplitudes in \eref{4p1} and \eref{4p2}, it is straightforward to recognize that
\bea
{\cal A}_{\rm NLSM^{+4}}(\vec{\pmb\sigma}_4)={s^2+t^2+u^2\over2}\,{\cal A}_{\rm NLSM}(\vec{\pmb\sigma}_4)\,.~~\label{4p-N+4}
\eea

\begin{figure}
  \centering
  % Requires \usepackage{graphicx}
  \includegraphics[width=12cm]{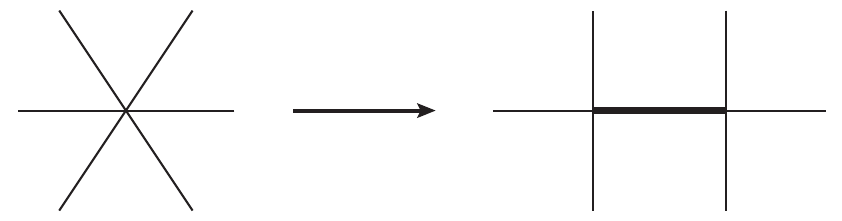} \\
  \caption{Split the $6$-point vertex to $4$-point ones. The bold line corresponds to the inserted propagator.}\label{split}
\end{figure}

With the $4$-point amplitudes determined in \eref{4p-N+4}, we can construct the general ${\rm NLSM}^{+4}$ amplitudes with arbitrary number of external pions from them. Each $4$-point ${\rm NLSM}^{+4}$ amplitude can be interpreted as the interaction between $4$ pions through the quadrivalent vertex which carries higher-derivative. For simplicity, let us call such quadrivalent vertex as the ${\rm NLSM}^{+4}$ vertex, and call the quadrivalent vertex correspond to the ordinary $4$-point NLSM amplitude as the NLSM vertex. In general, the ${\rm NLSM}^{+4}$ and NLSM amplitudes with arbitrary number of external pions, receive contributions from $2\ell$-point vertices with infinite tower of $\ell$. However, one can always split any higher-point vertex into $4$-point ones by inserting $1=D/D$, where $1/D$ are propagators which connect resulted quadrivalent vertices. An example for the decomposition of $6$-point vertex is shown in Figure.\ref{split}. Thus, we can consider only $4$-point vertices when discussing NLSM and ${\rm NLSM}^{+4}$ amplitudes. It is worth to emphasize that we do not fix the exact formula of any off-shell vertex, although the corresponding on-shell $4$-point amplitude is unique. In other words, we allow each vertex to have the freedom of taking different formulas, especially absorbing contributions from splitting higher-point vertices. This ambiguity has no influence on our construction for amplitudes, since in our method the numerators of amplitudes are determined by exploiting appropriate soft behaviors, without the aid of Feynman rules.

\begin{figure}
  \centering
  % Requires \usepackage{graphicx}
  \includegraphics[width=12cm]{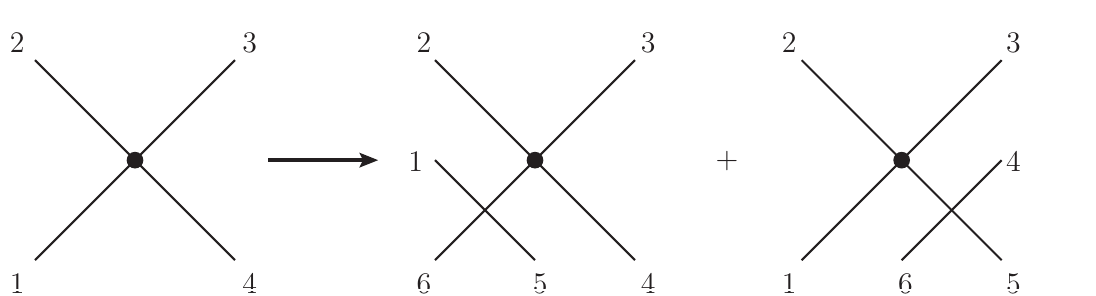} \\
  \caption{Inserting legs $5,6$ into the $4$-point $\rm NLSM^{+4}$ amplitude. The big bold point represents the $\rm NLSM^{+4}$ vertex.}\label{insert}
\end{figure}

The mass dimension implies that each $2n$-point ${\rm NLSM}^{+4}$ amplitude involves one quadrivalent ${\rm NLSM}^{+4}$ vertex and $(n-2)$ quadrivalent NLSM vertices. Thus, inserting new external pions into the amplitude ${\cal A}_{\rm NLSM^{+4}}(\vec{\pmb\sigma}_{2m})$ via a NLSM vertex gives a part of the full amplitude ${\cal A}_{\rm NLSM^{+4}}(\vec{\pmb\sigma}_{2m+2})$. In Figure.\ref{insert}, we show the Feynman diagrams correspond to inserting pions $5$ and $6$ into the $4$-point ${\rm NLSM}^{+4}$ amplitude. Similar as in the previous section, such insertion can be realized by inverting double soft theorem in \eref{doublesoft-theo} and \eref{doublesoft-fac}, which describes the double soft behavior at the $\tau^1$ order for $k_a\to\tau k_a$, $k_b\to\tau k_b$, $\tau\to 0$.
We now claim that any diagram in which $a$ and $b$ are connected to a ${\rm NLSM}^{+4}$ vertex has no contribution at the $\tau^1$ order. Suppose soft pions $a$, $b$ and another external pion $c$ are attached to a common ${\rm NLSM}^{+4}$ vertex, then the propagator $1/s_{abc}$ becomes on-shell in the limit $\tau\to0$, thus the full $n$-point amplitude factorizes into a $4$-point on-shell ${\rm NLSM}^{+4}$ amplitude and a $(n-2)$-point NLSM one. At the $4$-point ${\rm NLSM}^{+4}$ amplitude side, the leading order contribution arises from the unique on-shell amplitude in \eref{4p-N+4}. Such leading order is the $\tau^3$ order, as can be counted by using the expression in \eref{4p-N+4}, together with the constraints of momentum conservation and on-shell condition. Meanwhile, it is direct to find the leading orders for the propagator $1/s_{abc}$ and the $(n-2)$-point NLSM amplitude, which are $\tau^{-1}$ and $\tau^0$ respectively. Consequently, the leading order of contributions from diagrams those $a$ and $b$ are connected to ${\rm NLSM}^{+4}$ vertices is the $\tau^2$ order, thus are undetectable at the $\tau^1$ order under consideration.

Thus, at the $\tau^1$ order, the soft behavior ${\cal A}^{(1)_{ab}}_{\rm NLSM^{+4}}(\vec{\pmb\sigma}_{2m})$ only receives contributions from diagrams in which soft pions $a$ and $b$ are connected to a NLSM vertex. Therefore, ${\cal A}^{(1)_{ab}}_{\rm NLSM^{+4}}(\vec{\pmb\sigma}_{2m})$
satisfies the double soft theorem in \eref{doublesoft-theo}, factorizes into the soft factor in \eref{doublesoft-fac} and the $(n-2)$-pont subamplitude as follows
\bea
{\cal A}^{(1)_{ab}}_{\rm NLSM^{+4}}(\vec{\pmb\sigma}_{2m})=\big(S^{(1)_{ab}}_C+S^{(1)_{ab}}_D\big)\,{\cal A}_{\rm NLSM^{+4}}(\vec{\pmb\sigma}_{2m-2})\,.~~\label{doublesoft-2mp}
\eea
It is worth to emphasize that one can not expect the above factorization behavior if diagrams with $a$ and $b$ connected to ${\rm NLSM}^{+4}$ vertices also contribute to the $\tau^1$ order. The double soft behavior in \eref{doublesoft-2mp} will lead to a part of the full amplitude which is detectable at the subleading order. The complete amplitude can be determined by imposing appropriate symmetry, as will be seen in next subsection.

%%%%%%%%%%%%%%%%%%%%%
\subsection{higher-point ${\rm NLSM}^{+4}$ amplitudes}
\label{subsec-higher-point}
%%%%%%%%%%%%%%%%%%%%%%%%%%%%%%%%

Based on above discussions, we now construct the higher-point ${\rm NLSM}^{+4}$ amplitudes from $4$-point ones in \eref{4p-N+4}. Our goal is to expand any ${\rm NLSM}^{+4}$ amplitude to BAS ones, thus it is convenient to reformulate \eref{4p-N+4} as
\bea
{\cal A}_{\rm NLSM^{+4}}(\vec{\pmb\sigma}_4)&=&{\rm tr}(P_1\cdot P_2)\,{\cal A}_{{\rm N}\oplus{\rm B}}(1,2;\{3,4\}|\vec{\pmb\sigma}_4)
+{\rm tr}(P_3\cdot P_4)\,{\cal A}_{{\rm N}\oplus{\rm B}}(3,4;\{1,2\}|\vec{\pmb\sigma}_4)\nn
& &+{\rm tr}(P_1\cdot P_3)\,{\cal A}_{{\rm N}\oplus{\rm B}}(1,3;\{2,4\}|\vec{\pmb\sigma}_4)
+{\rm tr}(P_2\cdot P_4)\,{\cal A}_{{\rm N}\oplus{\rm B}}(2,4;\{1,3\}|\vec{\pmb\sigma}_4)\nn
& &+{\rm tr}(P_1\cdot P_4)\,{\cal A}_{{\rm N}\oplus{\rm B}}(1,4;\{2,3\}|\vec{\pmb\sigma}_4)
+{\rm tr}(P_2\cdot P_3)\,{\cal A}_{{\rm N}\oplus{\rm B}}(2,3;\{1,4\}|\vec{\pmb\sigma}_4)\nn
&=&\sum_{\vec{\pmb\a},|\pmb\a|=2}\,{\rm tr}(P_{\vec{\pmb\a}})\,{\cal A}_{{\rm N}\oplus{\rm B}}(\vec{\pmb\a};\{1,2,3,4\}\setminus\pmb\a|\vec{\pmb\sigma}_4)\,,~~\label{4p-N+4-reform}
\eea
where ${\rm tr}(T)\equiv T_\mu^{~\mu}$ for any tensor $T^{\mu\nu}$. To obtain \eref{4p-N+4-reform}, we used the observation that
each $4$-point amplitude is equivalent to $4$-point ${\rm NLSM}\oplus{\rm BAS}$ amplitudes with two external BAS scalars, and symmetrized the formula to manifest the permutation invariance. Start from \eref{4p-N+4-reform}, our method will lead to the expansion of general ${\rm NLSM}^{+4}$ amplitudes to ${\rm NLSM}\oplus{\rm BAS}$ ones. Then, by substituting the expansion in \eref{NB-BAS} iteratively, one can expand any ${\rm NLSM}^{+4}$ amplitude to BAS basis.

Now we construct the $6$-point ${\rm NLSM}^{+4}$ amplitude ${\cal A}_{\rm NLSM^{+4}}(\vec{\pmb\sigma}_6)$, with external legs encoded as $\{1,\cdots,6\}$.
Taking $k_5\to\tau k_5$, $k_6\to\tau k_6$, $\tau\to 0$, the double soft theorem requires
\bea
{\cal A}^{(1)_{56}}_{\rm NLSM^{+4}}(\vec{\pmb\sigma}_6)&=&\big(S^{(1)_{56}}_D+S^{(1)_{56}}_C\big)\,{\cal A}_{\rm NLSM^{+4}}(\vec{\pmb\sigma}_6\setminus\{5,6\})\nn
&=&\big(S^{(1)_{56}}_D+S^{(1)_{56}}_C\big)\,\Big(\sum_{\substack{\vec{\pmb\a},|\pmb\a|=2\\ \pmb\a\subset\{1,2,3,4\}}}\,{\rm tr}(P_{\vec{\pmb\a}})\,{\cal A}_{{\rm N}\oplus{\rm B}}(\vec{\pmb\a};\{1,2,3,4\}\setminus\pmb\a|\vec{\pmb\sigma}_6\setminus\{5,6\})\Big)\nn
&=&\sum_{\substack{\vec{\pmb\a},|\pmb\a|=2\\ \pmb\a\subset\{1,2,3,4\}}}\,{\rm tr}(P_{\vec{\pmb\a}})\,\Big(\big(S^{(1)_{56}}_D+S^{(1)_{56}}_C\big)\,{\cal A}_{{\rm N}\oplus{\rm B}}(\vec{\pmb\a};\{1,2,3,4\}\setminus\pmb\a|\vec{\pmb\sigma}_6\setminus\{5,6\})\Big)\nn
& &+\sum_{\substack{\vec{\pmb\a},|\pmb\a|=2\\ \pmb\a\subset\{1,2,3,4\}}}\,\Big(S^{(1)_{56}}_D\,{\rm tr}(P_{\vec{\pmb\a}})\Big)\,{\cal A}_{{\rm N}\oplus{\rm B}}(\vec{\pmb\a};\{1,2,3,4\}\setminus\pmb\a|\vec{\pmb\sigma}_6\setminus\{5,6\})\nn
&=&\sum_{\substack{\vec{\pmb\a},|\pmb\a|=2\\ \pmb\a\subset\{1,2,3,4\}}}\,{\rm tr}(P_{\vec{\pmb\a}})\,{\cal A}^{(1)_{5,6}}_{{\rm N}\oplus{\rm B}}(\vec{\pmb\a};\{1,2,3,4,5,6\}\setminus\pmb\a|\vec{\pmb\sigma}_6)\nn
& &+\tau^4\,\sum_{\substack{\vec{\pmb\a},|\pmb\a|=2\\ \pmb\a\subset\{1,2,3,4\}}}\,{\rm tr}(P_{\a_1,5\shuffle 6,\a_2})\,{\cal A}^{(0)_{5,6}}_{{\rm N}\oplus{\rm B}}(\a_1,5\shuffle 6,\a_2;\{1,2,3,4\}\setminus\pmb\a|\vec{\pmb\sigma}_6)\nn
& &+\tau^4\,\sum_{\substack{\vec{\pmb\a},|\pmb\a|=2\\ \pmb\a\subset\{1,2,3,4\}}}\,{\rm tr}(P_{\a_1,\a_2,5\shuffle 6})\,{\cal A}^{(0)_{5,6}}_{{\rm N}\oplus{\rm B}}(\a_1,\a_2,5\shuffle 6;\{1,2,3,4\}\setminus\pmb\a|\vec{\pmb\sigma}_6)\,,~~\label{6p-soft-N+4}
\eea
where elements in each length-$2$ ordered set $\vec{\pmb\a}$ are denoted as $\vec{\pmb\a}=\Spaa{\a_1,\a_2}$. In the above, the evaluation is paralleled to that
from \eref{n+4p-soft} to \eref{B2}, thus some details are omitted. The soft behavior in \eref{6p-soft-N+4} indicates
\bea
{\cal P}^{56}_{\rm NLSM^{+4}}(\vec{\pmb\sigma}_6)&=&\sum_{\substack{\vec{\pmb\a},|\pmb\a|=2\\ \pmb\a\subset\{1,2,3,4\}}}\,{\rm tr}(P_{\vec{\pmb\a}})\,{\cal A}_{{\rm N}\oplus{\rm B}}(\vec{\pmb\a};\{1,2,3,4,5,6\}\setminus\pmb\a|\vec{\pmb\sigma}_6)\nn
& &+\sum_{\substack{\vec{\pmb\a},|\pmb\a|=2\\ \pmb\a\subset\{1,2,3,4\}}}\,{\rm tr}(P_{\a_1,5\shuffle 6,\a_2})\,{\cal A}_{{\rm N}\oplus{\rm B}}(\a_1,5\shuffle 6,\a_2;\{1,2,3,4\}\setminus\pmb\a|\vec{\pmb\sigma}_6)\nn
& &+\sum_{\substack{\vec{\pmb\a},|\pmb\a|=2\\ \pmb\a\subset\{1,2,3,4\}}}\,{\rm tr}(P_{\a_1,\a_2,5\shuffle 6})\,{\cal A}_{{\rm N}\oplus{\rm B}}(\a_1,\a_2,5\shuffle 6;\{1,2,3,4\}\setminus\pmb\a|\vec{\pmb\sigma}_6)\,,~~\label{6p-part}
\eea
which serves as the detectable part when considering $k_5\to\tau k_5$, $k_6\to\tau k_6$, $\tau\to0$.

To achieve the complete amplitude, we break the adjacency of $5$ and $6$, and impose the invariance under the cyclic permutation of external legs. Such cyclic invariance is very natural since it states the basic character of ordering $\vec{\pmb\sigma}_6$. The above manipulation turns \eref{6p-part}
to
\bea
{\cal A}_{\rm NLSM^{+4}}(\vec{\pmb\sigma}_6)&=&\sum_{\vec{\pmb\a},|\pmb\a|=2,4}\,{\rm tr}(P_{\vec{\pmb\a}})\,{\cal A}_{{\rm N}\oplus{\rm B}}(\vec{\pmb\a};\{1,2,3,4,5,6\}\setminus\pmb\a|\vec{\pmb\sigma}_6)\,.~~~\label{6p-N+4-amp}
\eea
We interpret the above ${\cal A}_{\rm NLSM^{+4}}(\vec{\pmb\sigma}_6)$ as the full amplitude, due to the following reason. Each diagram contributes to ${\cal A}_{\rm NLSM^{+4}}(\vec{\pmb\sigma}_6)$ contains one NLSM vertex. Thus, for any particular diagram, one can always find two external pions $a'$ and $b'$ so that the contribution from this diagram is detectable when $a'$ and $b'$ are taken to be soft. In other words, all terms will be detected by considering all possible pairs $a',b'$. Traversing all pairs $a',b'$ is equivalent to symmetrizing the formula in \eref{6p-part}, thus we conclude that the expansion in \eref{6p-N+4-amp} is the correct answer.

\begin{figure}
  \centering
  % Requires \usepackage{graphicx}
  \includegraphics[width=8cm]{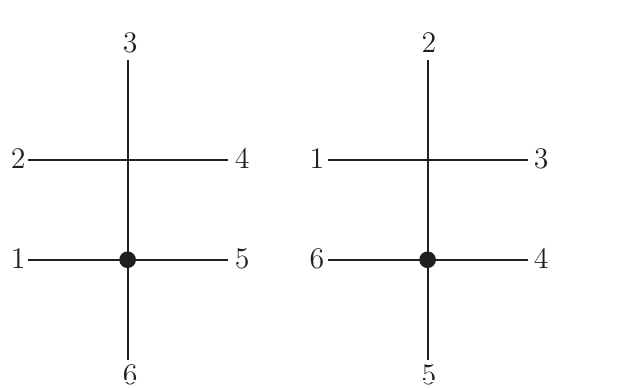} \\
  \caption{Graphs correspond to missed terms.}\label{miss}
\end{figure}

As mentioned before, Feynman diagrams in which soft pions are attached to ${\rm NLSM}^{+4}$ vertices do not contribute to the $\tau^1$ order under consideration. Let us figure out the contribution from such diagrams. Comparing the full amplitude \eref{6p-N+4-amp} and the detectable part \eref{6p-part}, one can find that the missed part includes the following terms
\bea
{\cal U}^{56}_{\rm NLSM^{+4}}(\vec{\pmb\sigma}_6)&=&\sum_{i\in\{1,2,3,4\}}\,{\rm tr}(P_{\Spaa{5,i}})\,{\cal A}_{{\rm N}\oplus{\rm B}}(i,5;6\cup\{\{1,2,3,4\}\setminus i\}|\vec{\pmb\sigma}_6)\nn
& &+\sum_{i\in\{1,2,3,4\}}\,{\rm tr}(P_{\Spaa{6,i}})\,{\cal A}_{{\rm N}\oplus{\rm B}}(i,6;5\cup\{\{1,2,3,4\}\setminus i\}|\vec{\pmb\sigma}_6)\nn
& &+{\rm tr}(P_{\Spaa{6,5}})\,{\cal A}_{{\rm N}\oplus{\rm B}}(5,6;\{1,2,3,4\}|\vec{\pmb\sigma}_6)\,.~~\label{6p-undete}
\eea
In the above, the first and third terms vanish at the $\tau^1$ order when taking only the pion $5$ to be soft, while the second and third terms vanish at the $\tau^1$ order when taking only $6$ to be soft, since the leading soft factor for BAS scalars \eref{soft-fac-s-0} vanishes for amplitudes with only two external BAS scalars. Because of the vanishing at $\tau^1$ order in the single soft limit, terms in \eref{6p-undete} can not be restored by only breaking the adjacency of $5$ and $6$. Thus we interpret ${\cal U}^{56}_{\rm NLSM^{+4}}(\vec{\pmb\sigma}_6)$ in \eref{6p-undete} as the contribution from diagrams those $5$ and $6$ are connected to ${\rm NLSM}^{+4}$ vertices, as shown in Figure.\ref{miss}. Diagrams in Figure.\ref{miss} are related to two diagrams at the r.h.s of Figure.\ref{insert} via the cyclic permutation among external legs, however the contributions from them do not inherit such symmetry. This is because of in our method we do not fix the exact formula of numerators of off-shell vertices, as discussed previously. In particular, since all higher-point vertices can be decomposed into quadrivalent ones, one can always split a $6$-point ${\rm NLSM}^{+4}$ vertex into a $4$-point NLSM vertex and another ${\rm NLSM}^{+4}$ vertex, make the corresponding contribution to be detectable. This freedom of decomposition breaks the symmetry among diagrams in Figure.\ref{insert} and Figure.\ref{miss}.

Repeating the recursive construction for obtaining $6$-point amplitudes in \eref{6p-N+4-amp}, the general ${\rm NLSM}^{+4}$ amplitudes with $2m$ external pions can be found as
\bea
{\cal A}_{\rm NLSM^{+4}}(\vec{\pmb\sigma}_{2m})&=&\sum_{\substack{\vec{\pmb\a},|\pmb\a|=2k,\\1\leq k\leq m-1}}\,{\rm tr}(P_{\vec{\pmb\a}})\,{\cal A}_{{\rm N}\oplus{\rm B}}(\vec{\pmb\a};\pmb{p}_{2m}\setminus\pmb\a|\vec{\pmb\sigma}_{2m})\,.~~\label{NLSM+4-2mp}
\eea
We have verified the correct factorizations of the above formula, until $10$-point.
We also verified that the amplitudes in \eref{NLSM+4-2mp} contain terms those various propagators are canceled by numerators,
therefore receive contribution from a variety of higher-point interactions. Since an amplitude represented by \eref{NLSM+4-2mp} is constructed from \eref{4p-N+4} by inserting pions via ordinary NLSM amplitudes, it can naturally be understood as the NLSM$^{+4}$ amplitude with a single insertion of local NLSM$^{+4}$ operator. Such single NLSM$^{+4}$ operator
serves as the leading higher-derivative correction to pure NLSM amplitudes, while NLSM ones are known as the leading contribution to the full pion scattering. Based on our method, the uniqueness of these subleading pion amplitudes \eref{NLSM+4-2mp} are ensured by the uniqueness of on-shell $4$-point ${\cal D}+4$ solution in \eref{4p-N+4}, albeit the off-shell forms of NLSM$^{+4}$ vertices have not been fixed.

In our recent work \cite{Zhou:2024qwm}, the YM amplitudes with single insertion of $F^3$ operator are found to be
\bea
{\cal A}_{\rm YM^{+2}}(\vec{\pmb{\sigma}}_n)=\sum_{\vec{\pmb{\rho}}}\,{\rm tr}(F_{\vec{\pmb{\rho}}})\,
{\cal A}_{\rm YMS}(\vec{\pmb{\rho}};\pmb{g}_n\setminus\pmb\rho|\vec{\pmb{\sigma}}_n)\,,~~\label{YM+2}
\eea
where
\bea
F^{\mu\nu}_{\vec{\pmb{\rho}}}=(f_{\rho_k}\cdots f_{\rho_1})^{\mu\nu}\,,~~~~{\rm for}~\vec{\pmb{\rho}}=\{\rho_1,\cdots,\rho_k\}\,.
\eea
When $n=2m$, we observe that the transmutation operator ${\cal L}_{2m}$ in \eref{Defin-L} transmutes the YM$^{+2}$ amplitudes in \eref{YM+2}
to NLSM$^{+4}$ ones as
\bea
{\cal A}_{\rm NLSM^{+4}}(\vec{\pmb\sigma}_{2m})={\cal L}_{2m}\,{\cal A}_{\rm YM^{+2}}(\vec{\pmb{\sigma}}_{2m})\,.~~\label{trans-YM+2}
\eea
%

%%%%%%%%%%%%%%%%%%%%%%%%%%%%%%%
\section{Discussion}
\label{sec-summary}
%%%%%%%%%%%%%%%%%%%%%%%%%%%%%%%

In this paper, a new approach for constructing tree amplitudes of scalar EFTs is proposed. Based on the phenomenon that pion amplitudes
only have $2\ell$-point interactions with $\ell$ the positive integers, it is natural to exploit the double soft behavior to build amplitudes.
We first bootstrap the $4$-point NLSM amplitudes, and reinterpret them as ${\rm NLSM}\oplus{\rm BAS}$ amplitudes with
$2$ external pions and $2$ external BAS scalars, then construct ${\rm NLSM}\oplus{\rm BAS}$ amplitudes with more external BAS scalars by inverting
the leading soft theorem for BAS amplitudes. The double soft theorem for pions can be determined by employing the resulted
$(n+2)$-point ${\rm NLSM}\oplus{\rm BAS}$ amplitudes. Then, based on the assumption of the universality of soft behavior, the general
${\rm NLSM}\oplus{\rm BAS}$ and NLSM amplitudes are determined by inverting double soft theorem for pions. Along this line, the pion
amplitudes with leading higher-derivative correction, which contain the single insertion of local NLSM$^{+4}$ operator, are also constructed.

Our method naturally represents the resulted amplitudes as universal expansions to ${\rm NLSM}\oplus{\rm BAS}$ amplitudes. Such expansions
ultimately yield the expansions to pure BAS basis. From such expansions, we observe that the BCJ relations, which are regarded as the central
physical constraint in \cite{Brown:2023srz}, are satisfied automatically. More explicitly, our results indicates the expansion
\bea
{\cal A}(\vec{\pmb\sigma}_n)=\sum_{\vec{\pmb\sigma}'_n}\,{\cal C}(\vec{\pmb\sigma}'_n)\,{\cal A}_{\rm BAS}(\vec{\pmb\sigma}'_n|\vec{\pmb\sigma}_n)\,,
\eea
where coefficients ${\cal C}(\vec{\pmb\sigma}'_n)$ are independent of the ordering $\vec{\pmb\sigma}_n$.
In the above, amplitudes ${\cal A}(\vec{\pmb\sigma}_n)$ can be understood as NLSM$\oplus$BAS amplitudes, NLSM amplitudes,
as well as NLSM$^{+4}$ amplitudes, which carry the ordering $\vec{\pmb\sigma}_n$ among all external legs. Since the BAS amplitudes
with different $\vec{\pmb\sigma}_n$ satisfy BCJ relations, amplitudes ${\cal A}(\vec{\pmb\sigma}_n)$ automatically inherit such property.

We end with a brief discussion on constructing pion amplitudes with higher mass dimension.
As discussed in section \ref{subsec-4p-higher}, any higher-point interaction can be decomposed
into quadrivalent ones, thus the construction involves two levels. The first one, the basic $4$-point interactions are still
NLSM and NLSM$^{+4}$ ones. At this level, we can construct the $2(h+1)$-point NLSM$^{+4h}$ amplitudes first, these amplitudes contain only
NLSM$^{+4}$ quadrivalent vertices. Then the general NLSM$^{+4h}$ amplitudes can be achieved by inverting double soft theorems in \eref{doublesoft-theo}
and \eref{doublesoft-fac}.
To get the desired $2(h+1)$-point NLSM$^{+4h}$ amplitudes, one natural way is to find the double soft theorem for pions attached to a
NLSM$^{+4}$ vertex (at $\tau^2$ or higher order), then invert such double soft theorem to construct higher-point NLSM$^{+4h}$ amplitudes recursively. The derivation for
this new double soft theorem requires additional new technic. Since the same technic can be applied to treat the YM amplitudes with higher-derivative interactions, we will do this work elsewhere. The second level is that the basic $4$-point amplitudes have mass dimension
higher than ${\cal D}+4$.
Along the line in this paper, the challenge is to
solve the lowest $4$-point amplitudes. In \cite{Brown:2023srz}, the authors found that the uniqueness of solution is maintained up to $h=5$, for general ${\rm NLSM}^{+2h}$
amplitudes. Starting from $h=6$, there are two or more independent solutions. Thus, we expect the approach in this paper to be effective up
to $h=5$. The study on this interesting topic is also left to the future work.

%%%%%%%%%%%%%%%%%%%%%%%
\section*{Acknowledgments}
%%%%%%%%%%%%%%%%%%%%%%

This work is supported by NSFC under Grant No. 11805163.

%%%%%%%%%%%%%%%%%%%%%%%%%%%%%%%%%%%

\bibliographystyle{JHEP}

\bibliography{reference}

\end{document}